\documentclass[aps,prd,reprint,twocolumn,groupedaddress,superscriptaddress,nofootinbib,longbibliography,showpacs,showkeys]{revtex4-1}
\usepackage{graphics}%
\usepackage[mathcal]{euscript}
\usepackage[latin1]{inputenc}
\usepackage{latexsym}
\usepackage{hyperref}
\usepackage{amsmath}    
\usepackage{graphicx}   
\usepackage{verbatim}  
\usepackage{color}      
\usepackage{subfigure}  
\usepackage{hyperref}   
\usepackage{bm}
\usepackage{graphicx}
\usepackage{amssymb}
\usepackage{bbm}
\usepackage{float}
\usepackage{slashed}
\usepackage{amsfonts}
\usepackage{euscript}
\usepackage{amsmath}    
\usepackage{mathrsfs} 
\usepackage{bbding}
\usepackage{pifont}
\usepackage{cancel}
\usepackage{multirow}
\usepackage{soul}

\usepackage[usenames,dvipsnames]{xcolor}
\hypersetup{colorlinks=true, citecolor=Bittersweet, linkcolor=Bittersweet,
urlcolor=Bittersweet}

\newcommand{\cmark}{\ding{51}}%
\newcommand{\xmark}{\ding{55}}%

\begin{document}%

\title{Phenomenological aspects of black holes beyond general relativity}


\author{Ra\'ul Carballo-Rubio}
\email{raul.carballorubio@sissa.it}
\affiliation{SISSA, International School for Advanced Studies, Via Bonomea 265, 34136 Trieste, Italy}
\affiliation{INFN Sezione di Trieste, Via Valerio 2, 34127 Trieste, Italy}
\author{Francesco Di Filippo}
\email{francesco.difilippo@sissa.it}
\affiliation{SISSA, International School for Advanced Studies, Via Bonomea 265, 34136 Trieste, Italy}
\affiliation{INFN Sezione di Trieste, Via Valerio 2, 34127 Trieste, Italy}
\author{Stefano Liberati}
\email{liberati@sissa.it}
\affiliation{SISSA, International School for Advanced Studies, Via Bonomea 265, 34136 Trieste, Italy}
\affiliation{INFN Sezione di Trieste, Via Valerio 2, 34127 Trieste, Italy}
\author{Matt Visser}
\email{matt.visser@sms.vuw.ac.nz}
\affiliation{School of Mathematics and Statistics, Victoria University of Wellington; PO Box 600, Wellington 6140, New Zealand}

\bigskip
\begin{abstract}

While singularities are inevitable in the classical theory of general relativity, it is commonly believed that they will not be present when quantum gravity effects are taken into account in a consistent framework. In particular, the structure of black holes should be modified in frameworks beyond general relativity that aim at regularizing singularities. Being agnostic on the nature of such theory, in this paper we classify the possible alternatives to classical black holes and provide a minimal set of phenomenological parameters that describe their characteristic features. The introduction of these parameters allows us to study, in a largely model-independent manner and taking into account all the relevant physics, the phenomenology associated with these quantum-modified black holes. We perform an extensive analysis of different observational channels and obtain the most accurate characterization of the viable constraints that can be placed using current data. Aside from facilitating a critical revision of previous work, this analysis also allows us to highlight how different channels are capable of probing certain features but are oblivious to others, and pinpoint the theoretical aspects that should be addressed in order to strengthen these tests.

\end{abstract}

\maketitle

\def\HRULE{{\bigskip\hrule\bigskip}}

\section{Introduction}

Black holes comprise a remarkably elegant collection of solutions of the classical Einstein field equations. Aside from their rich mathematical structure, they are nowadays accepted as legitimate astrophysical objects and are routinely used in order to explain astrophysical observations  (e.g., \cite{Celotti1999,Narayan2005}). Moreover, they provide essentially the best laboratory (only rivaled by cosmology) in which strong gravity, and perhaps even quantum gravity, can be put to test. This makes black hole physics a topic of interest for a wide range of physicists, working on theoretical as well as observational aspects.

While this provides a fertile soil for the growth of strong interactions between theorists and experimentalists, these interactions are sometimes not as coherent as they might be, and many aspects of them can be certainly improved. On the one hand, observational studies often disregard the precise meaning of certain mathematical definitions, to the extent of discussing the observability of notions that are by definition not observable (the best example is the notion of event horizon~\cite{Visser2014}). On the other hand, theoretical studies tend to oversimplify the connection between theoretical models and observations, sometimes lacking a careful analysis of what is realistically observable or not.

The discussion in this paper is divided in two sections. Sec.~\ref{sec:taxonomy} consists of a brief review of the different scenarios that have been proposed in the literature in order to avoid theoretical problems that are inherent to classical and semiclassical black holes (e.g., the singularity or information loss problems), including a discussion about the plausible connections between some of these scenarios. Sec. \ref{sec:channels} focuses on the observational tests that are available to probe the structure of astrophysical black holes, with the goal of emphasizing how close we are to testing the existence of all the relevant mathematical elements of these objects as described by general relativity, and parametrizing the room that is still available for alternatives in terms of a set of phenomenological parameters introduced here for the first time. The meaning of these phenomenological parameters is illustrated by comparison with the scenarios described in Sec. \ref{sec:taxonomy}.

This work is not intended to be an exhaustive review that encompasses all the previous developments in the subject. Rather, our goal is providing a critical and thorough assessment of the observational and theoretical uncertainties in our current understanding of astrophysical black holes. Hence, the selection of topics in this paper does not follow from a completeness criterion. We have selected instead a number of issues that remained obscure or that demanded a more detailed analysis, and we provide the necessary background to understand their importance and clarify them. Sec. \ref{sec:taxonomy} provides the background and motivation for the phenomenological parametrization at the core of this paper. This parametrization, introduced at the beginning of Sec. \ref{sec:channels}, simplifies extracting the phenomenology of different theoretical proposals and also permits us to pinpoint the crucial elements in the subsequent phenomenological discussion. Particular attention is devoted to certain phenomenological aspects that have been overlooked in the literature to date, and which are important enough to modify strongly the main conclusions in previous works.

\section{A taxonomy of proposals \label{sec:taxonomy}}

\subsection{Regular black holes \label{sec:reg}}

An ubiquitous property of black hole solutions is the appearance of singularities. It is clear that physics beyond classical general relativity is needed in order to avoid this singular behavior. Regular black holes represent an attempt to yield geometries that are almost practically identical to the black hole solutions of general relativity, but that deviate from the latter significantly at the core of black holes. These deviations from the classical behavior are justified as an effective description of the necessary (yet unknown) physics to produce a non-singular behavior. 

In the past, the main motivation behind the study of these geometries has been the extraction of new predictions regarding the endpoint of the evaporation process and the possible resolution of the so-called information loss problem. It is perhaps worth stressing that, while regular black hole scenarios can ameliorate this problem due to the absence of singularities, they do not automatically provide a full resolution of the latter. Indeed, this also depends on features of the evaporation process that are not fixed without additional assumptions. While not everyone is uncomfortable with information loss~\cite{Unruh2017}, undeniably this problem is still a strong motivation for theoretical research in the field~\cite{Polchinski2016,Marolf2017}, albeit we shall not delve on this issue here.

Regular black holes constitute a static alternative to black holes, meaning that this approach is agnostic to the dynamical processes that lead the formation of such objects and it only considers the regularized static geometry (dynamics is added in a second stage only through Hawking evaporation). In the absence of other guiding principles, one prescribes the following two postulates~\cite{Hayward2005,Ansoldi2008,Frolov2014} (see \cite{Olmo2015,Olmo2015b,Olmo2016,Olmo2017,Bejarano2017,Menchon2017,Cano2018} for alternative perspectives):
\begin{itemize}
\item{It is possible to replace the singular core of classical black holes by a smooth spacetime region in which the metric does not necessarily satisfy the vacuum Einstein equations.}
\item{This effective description contains no singularities. In particular, physical observables such as curvature invariants are bounded everywhere.}
\end{itemize}

These two postulates lead to an effective geometric description which, in the static and spherically symmetric case (which we focus on for simplicity; note also that we are working in four dimensions), takes the form~\cite{Bardeen1968,Borde1996,Dymnikova1992,Frolov2017}
\begin{equation}
\text{d}s^2=-e^{-2\phi(r)}F(r)\text{d}t^2+\frac{\text{d}r^2}{F(r)}+r^2\text{d}\Omega^2,\label{eq:hw1}
\end{equation}
where $\text{d}\Omega^2=\text{d}\theta^2+\sin^2\theta\,\text{d}\varphi^2$ and
\begin{equation}
F(r)=1-\frac{2m(r)}{r}.\label{eq:hw2a}
\end{equation}
This geometry reduces to Schwarzschild for  
\begin{equation}
m(r)=M\in\mathbb{R},\qquad \phi(r)=\Phi\in\mathbb{R}. 
\end{equation}
As a specific example, let us consider a  ``minimal model'' introduced in~\cite{Hayward2005}, given by,
\begin{equation}
F(r)=1-\frac{r_{\rm s}}{r+\ell_\star^3/r^2},\qquad \phi(r)=\Phi\in\mathbb{R}.\label{eq:hw2}
\end{equation}
Here $r_{\rm s}=2M$, $\ell_\star=(r_{\rm s}\ell_{\rm P}^2)^{1/3}$, with $\ell_{\rm P}$ the Planck length. If we set $\ell_{\rm P}=0$ (equivalently, $\ell_\star=0$), the Schwarzschild geometry is recovered. For $\ell_{\rm P}\neq0$, this geometry interpolates from Schwarzschild wherever $\ell_\star/r\ll1$, to de Sitter with cosmological constant $\Lambda=3\ell^2$ wherever $\ell_\star/r\gg1$. 

In spite of its simplicity, the above model exhibits a series of features shared by basically all the proposals for regular black holes in the literature, namely: (i) there is an additional parameter $\ell_\star\ll r_{\rm s}$, the value of which must be specified, and that is used to measure the size of the interior region in which deviations from the classical geometry become important, (ii) the geometry in the core is typically de Sitter, a feature that was previously and independently associated with the behavior of matter at high densities~\cite{Sakharov1966,Gliner1966}, and (iii) regularity at $r=0$ plus asymptotic flatness imply the presence of even numbers of zeroes for the function $F(r)$ and hence of at least two horizons (see, e.g.,~\cite{Carballo-Rubio2018r}). In all the cases considered in this article, this implies that the outer trapping horizon is also associated to an inner horizon.

The above discussion should suffice to explain why regular black holes are often regarded as a minimal extension of their classical counterparts. Indeed, the features described  above are arguably quite conservative, and imply very weak (or possibly entirely absent) deviations from classical black holes outside of the trapping horizon (we shall further discuss this point in what follows). Let us then investigate the possible observational effects that this class of mimickers could yield.

These are rather limited if the model breaks down at a late stage of the evaporation process, where significant deviations from classical black hole dynamics are expected to arise. The only possibility is to consider the deviations from the Schwarzschild geometry in the near-horizon region or even outside the black hole, and check how these modifications affect physical processes (e.g.,~\cite{Amir2015,Amir2016,Gwak2017}) that take place in these regions during periods of time which are much smaller than the typical timescale for the dynamical evolution of the regular black hole itself. From Eq.~\eqref{eq:hw2}, it follows that for $r/\ell_\star\gg1$, the modifications from the Schwarzschild geometry are suppressed as
\begin{equation}
F(r\gg\ell_\star)=\mathscr{O}\left(\frac{r_{\rm s}}{r}\times\frac{\ell_\star^3}{r^3}\right).
\end{equation}
In terms of curvature invariants, these modifications are extremely tiny for typical values of $\ell_\star$, which makes them extremely difficult to measure observationally. Moreover, it is worth stressing that there is no good reason to expect these modifications to be physical. The particular regularization in Eq.~\eqref{eq:hw2} uses a function $\ell^3_\star/r^3$ with a decaying but nonzero tail, which is not a feature that is necessarily physical, but can be an artifact of this simple example. It is possible (and arguably, more compatible with the basic ideas about quantum gravity motivating regular black hole scenarios) to use functions of compact support to regularize the singular classical geometries. 

Just to give a simple example, let us consider the bump function $\mathcal{B}(x)$\\
\begin{equation}
\mathcal{B}(x)=\left\{\begin{array}{ll}e^{x^2/(x^2-1)},&\qquad x\in(-1,+1);\\0, &\qquad |x|\in[1,\infty).\\\end{array}\right.
\end{equation}
This function is both infinitely smooth ($C^\infty$) and has compact support on $x\in[-1,1]$, although it is not analytic ($C^\omega$) for $x=\pm1$. It has also been normalized such that $\mathcal{B}(0)=1$. Using this function we can define
\begin{equation}
F(r)=1-\frac{r_{\rm s}}{r+\mathcal{B}(r/\ell_\star)\ell_\star^3/r^2}.
\end{equation}
This prescription is physically just as well motivated as that in Eq.~\eqref{eq:hw2}, but the corresponding geometry is now exactly Schwarzschild for $r\geq\ell_\star$. As long as $\ell_\star<r_{\rm 
s}$, there is no possible way to detect the effects associated with the regularization of the core using external observational probes (which are the ones we have at our disposal).

This comment applies to regular black holes in general: their static or stationary geometries can be prescribed to  exactly  match the corresponding classical geometries outside the regular core and, moreover, there is no physical reason why this equivalence cannot hold. This would make these geometries observationally indistinguishable for physical processes in which the approximation of considering static or stationary spacetimes is reasonable.

Indeed, given the above discussion it might appear that regular black holes are perfect mimickers in the sense that they can conjugate a conservative local regularization of  the interior singularity with an exterior geometry which is virtually undistinguishable from that of a classical black hole. 
However, it has been recently shown that these scenarios are inherently inconsistent~\cite{Carballo-Rubio2018r}.

The basic problem is twofold: inner horizons are unstable; whereas the end-point of the Hawking evaporation process leads to near-extremal horizons, with divergent evaporation timescales. In other words, these models fail to be self-consistent, and only new ingredients may lead to a satisfactory and complete picture of black hole evaporation. However, the nature of these necessary new assumptions is not known at the moment. Indeed, the results of~\cite{Carballo-Rubio2018r} may be taken as a rather strong indication that postulating short-range regularizations of black hole spacetimes could be outright inconsistent. This may be taken as a hint of a no-go theorem that is worth exploring.

We can then summarize the situation with regular black holes as follows:
\begin{itemize}
\item{There might be tiny corrections to the classical geometries in regions which are observationally available, taking the form of polynomial tails in the dimensionless variable $\ell_\star/r$. However, the existence of these tails is questionable.}
\item{Known models of regular black holes, satisfying the two postulates above, cannot be consistent through all the stages of evaporation and are generically unstable. The additional information that is needed to supply in order to describe consistently the evolution of the system adds large uncertainties regarding the fate of these objects. First of all, one has to decide the point in time at which the new effects come into play. Furthermore, it is not clear how the evolution would continue after this happens.}
\end{itemize}

To conclude this section, let us mention that the complete dynamical evolution of regular black holes is much less understood that their static and stationary limits. Certain rough guidelines for the construction of the complete dynamical geometry were given in~\cite{Hayward2005}, but there are still certain important details to be filled. One of them is the backreaction of classical perturbations and possibly Hawking radiation on the inner trapping horizon. This has been partially considered in~\cite{Frolov2016,Frolov2017}, and is in fact one of the main ingredients leading to the inconsistency of these models~\cite{Carballo-Rubio2018r}. 

The role that this instability plays in the evolution of regular black holes (and also classical black holes, where it is also present) is still to be studied. In particular, it may be the case that the backreaction on the inner trapping horizon is important enough to destabilize the overall structure of the regular black hole on timescales which are shorter than the naive evaporation time of the black hole. These effects might imply that black holes ``open up'' faster than expected. This possibility connects naturally with other scenarios proposed in the literature, which are described in the following two sections.

\subsection{Bouncing geometries \label{sec:vaid2}}

Instead of focusing on the description of the regular geometries that may describe black holes once formed, let us now analyze in more detail the fate of one of the main actors in the formation process: matter. Black holes are formed from collapsing lumps of matter. It is therefore of utmost importance to understand the physical mechanisms that may prevent these lumps of matter from collapsing completely, resulting in a singularity, and the implications that follow. 

In practical terms, we can divide the spacetime of a collapsing (spherically symmetric) distribution of matter in two regions: the external one, which is idealized as being vacuum, and the internal one which describes the interior of the collapsing matter. In the previous section we have been dealing mainly with the external geometry. However, the physics of the internal geometry is important as well, and in fact it might be the most important one regarding the regularization of the singularity \cite{Barcelo2015}.

The non-singular core of the regular black holes discussed above can be understood as the result of the tendency of matter towards collapsing, together with the existence of repulsive forces of quantum-mechanical nature that are triggered when Planckian curvatures are reached. However, from this perspective it seems unnatural that these two tendencies cancel exactly in order to stabilize the collapsing distribution of matter at a fixed radius no matter what the initial conditions are. Dynamically, it seems more natural to expect that the existence of a repulsive core would lead generally to bouncing geometries (while asymptotically stationary solutions with a small core may be reached for very particular values of the initial conditions).

This kind of bouncing behavior was first studied in~\cite{Frolov1979,Frolov1981,Hajicek2002,Ambrus2005}, but has been independently proposed more recently in scenarios inspired by emergent gravity~\cite{Barcelo2010,Barcelo2014e,Barcelo2014} and loop quantum gravity~\cite{Rovelli2014,Haggard2014}. Regardless of the special details of each particular implementation of this idea, there are some aspects that are shared by different proposals (see however~\cite{Ashtekar2005,Ashtekar2008,Corichi2015,Olmedo2017,Ashtekar2018,Ashtekar2018b,Bambi2013,Bambi2016,Bianchi2018} for alternative bouncing scenarios):
\begin{itemize}
\item{\emph{Timescale~\cite{Barcelo2010,Haggard2014}:} the existence of a bounce of matter would be physically meaningful only if the timescale for the bounce (suitably defined in terms of the proper time of the relevant observers) is shorter than the (naive) evaporation time of the black hole, 
\begin{equation}\label{eq:t3}
\tau_{\rm B}<\tau^{(3)}\sim t_{\rm P}(M/m_{\rm P})^3
\end{equation}
(if this condition is not satisfied, the black hole formed in the collapse would evaporate before the bounce could take place).
}
\item{\emph{Modifications of the near-horizon geometry~\cite{Barcelo2014e,Haggard2014}:} the external geometry of the spacetimes in which the bounce of the distribution of matter can be observed by external observers in the original asymptotic region must include modifications of the near-horizon geometry, even if the bounce of matter takes place much deeper inside the gravitational potential well (roughly, when the density of the radius of the collapsing structure is Planckian). The ultimate reason is that these geometries must interpolate between a black-hole geometry and a white-hole geometry. The only continuous way to define this interpolation involves modifications of the geometry up to a certain radius $r_\star>r_{\rm s}$, the value of which is typically constrained by $(r_\star-r_{\rm s})\lesssim r_{\rm s}$. This point can be illustrated using Painlev\'e-Gullstrand coordinates:
\begin{equation}
\text{d}s^2=-\text{d}t^2+[\text{d}r-fv(r)\text{d}t]^2+r^2\text{d}\Omega^2,
\end{equation}
where $f=\mp1$ correspond to a black hole or a white hole in these coordinates, respectively, and $v(r)=\sqrt{r_{\rm s}/r}$. In order to modify the geometry continuously from one case to the other, a function $f(t,r)$ that interpolates between these values, and which is nonzero at $r=r_{\rm s}$, is needed.
}
\end{itemize}
Plausible values for the timescale $\tau_{\rm B}$ are
\begin{equation}
\tau_{\rm B}=\tau^{(j)}\sim t_{\rm P}(M/m_{\rm P})^j,\qquad j=1,2.
\end{equation}
These two values verify $\tau^{(j)}\ll\tau^{(3)}$. For $j=2$ the bounce cannot be time-symmetric due to the unstable nature of white holes \cite{Barcelo2015b,DeLorenzo2015}. Let us note that these timescales are generally multiplied by logarithmic factors that have not been written explicitly but become relevant for certain values of the parameters involved. The shortest possible time for the bounce to take place is $\tau_{\rm B}=\tau^{(1)}$, as proposed originally in~\cite{Barcelo2010,Barcelo2014e,Barcelo2014}. Another possibility, conjectured in~\cite{Haggard2014,Christodoulou2016}, has $\tau_{\rm B}=\tau^{(2)}$. The second timescale is small if compared with $\tau^{(3)}$, but is still quite large so that, for a black hole to be exploding today, it must be a primordial black hole (though with different mass than in the case $\tau=\tau^{(3)}$ \cite{Hawking1974}). So far it does not seem possible to justify the timescale $\tau_{\rm B}=\tau^{(2)}$ through specific calculations such as the one in~\cite{Christodoulou2016}, while the timescale $\tau_{\rm B}=\tau^{(1)}$ arises in two independent calculations~\cite{Ambrus2005,Barcelo2016}.

Regarding possible observational channels, the second feature above (the modifications of the near-horizon geometry) might seem the most promising one at a first glance. Note that, in contraposition to regular black holes, these modifications are now a must. However, let us note that in the most natural scenarios, the modifications in these geometries are by construction $\mathscr{O}(1)$ only after the time $\tau_{\rm B}$. 

This implies that, if we are probing the geometry with physical processes during an interval of time $\Delta t$ (which could be, for instance, the time that an observed  photon took to go through this region of the geometry), any cumulative effects due to these deviations from the classical geometries would be suppressed by the dimensionless quotient $\Delta t/\tau_{\rm B}$. For $\tau_{\rm B}=\tau^{(2)}$ this dimensionless number will be generally small. Hence, it is reasonable to think that, unless we are considering physical processes with typical timescales comparable to the timescale of the bounce $\tau_{\rm B}$, it would be difficult to use these processes in order to learn about the geometry. It might be possible however that there exist non-trivial mechanisms that act to amplify the value of the number $\Delta t/\tau_{\rm B}$ when the latter is small, though at the moment this is just a conjecture. On the other hand, if one was able to perform experiments which monitor the evolution of the black hole during a time interval $\Delta t\sim \tau_{\rm B}$, the very end of the bounce process would be observable, which would certainly be more dramatic.

In consequence, in the absence of mechanisms that can compensate the small quotient $\Delta t/\tau_{\rm B}$, (looking for these amplification mechanisms would be an interesting exercise), the most promising observable feature of models with $\tau_{\rm B}=\tau^{(2)}$ is, again, the one associated with the end of the dynamical process; for regular black holes this was the evaporation process, but here is the very bounce of matter. The associated physics resides simply in the timescale $\tau_{\rm B}$, which for instance controls the typical size of the black holes for which these effects would be observable today.

Let us anyway assume that bounces with a timescale $\tau_{\rm B}$ take place in nature. One is then forced to answer the question: what happens after the bounce? There are essentially two possibilities. One possibility is that the bounce releases all the available energy and matter such that the black hole dissipates completely. This would imply that black holes disappear in a timescale $\tau_{\rm B}$, which can be compatible with observations only if $\tau_{\rm B}$ is large enough. This possibility is the one considered for instance in~\cite{Barrau2014,Barrau2015} and, as we have discussed, from a phenomenological perspective the most promising observational opportunity is the detection of this cataclysmic event (see Sec. \ref{sec:bursts} for a more detailed discussion). 

A second (and, arguably, more natural) possibility is that part of the energy content is released after the bounce, leaving a remnant that may again undergo renewed gravitational collapse~\cite{Barcelo2010,Barcelo2014,Barcelo2016}. In this scenario, it may be possible that a different kind of horizonless equilibrium configuration is reached, so that the bounce itself (which may occur several times) would be just a transient. This second possibility is arguably more rich observationally, as both the transient phase with timescale roughly $\tau_{\rm B}$ and the horizonless stable phase would have distinctive observational signatures which we shall now explore in the next section.

\subsection{Quasi black holes}

In order to encompass different alternatives in the literature, let us define a static and spherically symmetric quasi black hole in a rough way as a spacetime satisfying the following conditions: (i) the geometry is Schwarzschild above a given radius $R$ that is defined to be the radius of the object, (ii) the geometry for $r\leq R$ is not Schwarzschild, and (iii) there are no event or trapping horizons. In other words, this kind of geometry is qualitatively similar to that of a relativistic star, but with a typical radius of the structure $R$ that can be arbitrarily close to $r_{\rm s}$, hence violating the isotropic-pressure Buchdahl-Bondi bound~\cite{Buchdahl1959,Bondi1964} (let us note in passing that including pressure anisotropy permits one to attain more compact configurations that are not limited by the isotropic Buchdahl-Bondi bound~\cite{Cattoen2005}.) In fact, in this section we consider objects that can be characterized as having a surface. Configurations that fall within the definition above but do not have a surface (wormholes) are described in the next section.

There are several proposals in the literature for this kind of geometry (including gravastars~\cite{Mazur2004,Visser2003b,Mottola2010}, fuzzballs~\cite{Mathur2005,Guo2017}, and black stars~\cite{Barcelo2007,Barcelo2009}), but all of them present severe restrictions. In general terms, it is possible to prescribe (or derive from first principles~\cite{Carballo-Rubio2017}) this kind of geometry only when the outer geometry is that of non-rotating or slowly rotating black holes. Most importantly, there is virtually no knowledge about the dynamics of these objects; not only there are large gaps in the understanding of their possible formation mechanisms, but also of their behavior under other dynamical processes that they may undergo after formation. For instance, it is generally not clear how these objects interact with regular matter. There have been studies proposing Hawking radiation as the main ingredient to form these objects~\cite{Kawai2013,Kawai2014,Kawai2015,Baccetti2016,Baccetti2017,Baccetti2017b}, but these proposals share a number of problems (e.g.,~\cite{Chen2017,Juarez-Aubry2018}) that raise substantial doubts about their validity. It is also worth stressing that quasi black holes violate the assumptions of no-hair theorems (e.g., \cite{Cardoso2016b}), so that it is in principle possible that the external geometry is different than the Schwarzschild geometry in static (but not spherically symmetric) situations.

If quasi black holes are formed, this would require the existence of a transient before the system can settle down in this kind of configuration. The details of this transient are still largely unknown and would probably be rather complex, but in a first approach we can parametrize our ignorance in terms of another timescale, namely a relaxation timescale (in the bouncing scenario discussed above, this scale would be controlled by $\tau^{(1)}$, with logarithmic corrections depending on the values of certain parameters~\cite{Barcelo2015b}). This timescale could also show up in other events such as, for instance, the merger of two of these objects. As emphasized above, it is not known whether or not it is possible that the result of the merger of two quasi black holes is still a quasi black hole. However, if this is the case, it is reasonable to assume that the final quasi black hole state is reached after this relaxation timescale.

Aside from these transients, these objects are expected to lead to distinctive phenomenological signatures arising from interaction with light, matter and gravitational waves --- due to the large modifications of the geometry starting at $r=R$. In particular, while black holes are perfectly absorptive, quasi black holes do not necessarily satisfy this property (although there are again many uncertainties in this regard). Note, however, that any physical observable (i.e., a quantity that can be measured in an experiment) will go back continuously to its value for a black hole in the limit in which $\mu\rightarrow0$, where
\begin{equation}\label{eq:mudef}
\mu=1-\frac{r_{\rm s}}{R}.
\end{equation}
This observation would seem to imply that, for extremely small $\mu$, all observables should differ by very small amounts from the values they would take for a classical black hole (we can think about performing a Taylor expansion on the parameter $\mu$). This depends, however, on the functional dependence of observables on $\mu$. We will see that different observables display different behaviors (e.g., polynomial or logarithmic) with respect to $\mu$.

In order to develop some intuition on the typical values of $\mu$, let us make explicit the relation between $\mu$ and the distance between the surface and the would-be horizon. For $\mu\ll 1$, and if the surface is at a proper radial distance $\ell\ll r_{\rm s}$ from $r_{\rm s}$, one has
\begin{equation}\label{eq:muquad}
\mu\simeq\frac{1}{4}\left(\frac{\ell}{r_{\rm s}}\right)^2\simeq 7\times10^{-78}\left(\frac{M_\odot}{M}\right)^2\left(\frac{\ell}{\ell_{\rm P}}\right)^2.
\end{equation}
It is illustrative to consider for instance $\ell\sim\ell_{\rm P}$ and the mass corresponding to Sgr A*, $M=4\times 10^6\ M_\odot$, which yields $\mu\sim 10^{-91}$.

\subsection{Wormholes}

Wormholes are tunnels connecting different regions of spacetime and supported by large amounts of exotic matter or energy~\cite{Visser1995,Visser1989,Visser1989b,Dadhich2001,Visser2003,Kar2004}. The most interesting class of wormholes are the so-called traversable wormholes, that can be maintained open for enough time to allow geodesics to travel through them. Let us focus for simplicity on Morris-Thorne wormholes~\cite{Morris1988,Morris1988b}: time independent, non-rotating and spherically symmetric solutions of general relativity (with a suitable matter content) describing a bridge/passage between two asymptotically flat regions, not necessarily in the same universe. These objects are described by the metric
\begin{equation}\label{eq:whmetric}
\text{d}s^2=-e^{-2\phi(x)}\text{d}t^2+\text{d}x^2+r^2(x)\text{d}\Omega^2,
\end{equation}
where $x \in (-\infty,+\infty)$ and one requires the absence of event horizons and metric components that are at least $C^2$ in $x$. Asymptotic flatness for $x\to\pm\infty$ requires
\begin{equation}
\lim_{x\to\pm\infty}\frac{r(x)}{|x|}=1
\end{equation}
and
\begin{equation}
\lim_{x\to\pm\infty}\phi(x)=\Phi_\pm\in\mathbb{R}.
\end{equation}
On the other hand, the radius at the wormhole throat is $r_0=\mbox{min}\left\{r(x)\right\}$, which can always be chosen to be at $x=0$. The geometry \eqref{eq:whmetric} corresponds to flat spacetime, up to small corrections, far away from the throat. It is nevertheless possible to modify the external geometry so that it describes the gravitational field of a massive source, as in the Schwarzschild geometry~\cite{Visser1997,Cardoso2016a,Volkel2018}, or also to include rotation (e.g., \cite{Teo1998}) which would be eventually necessary in order to describe realistic astrophysical black holes. Moreover, as in the case of quasi black holes, the no-hair theorems do not directly apply to these objects, so that even in static situations it may be possible that higher multipoles take nonzero values.

This metric is a solution of the (non-vacuum) Einstein equations that requires a stress-energy tensor that violates the null energy condition, which states the positivity of the product $T_{\mu\nu} k^\mu k^\nu$ for any null-like vector $k^\mu$ (this implies that the weak, strong and dominant energy conditions are violated as well~\cite{Visser1995,Curiel2014}). Hence, the matter and energy content that keeps the throat open cannot have standard properties. These exotic properties may find a justification in the quantum properties of matter, when the latter is described in terms of quantum fields. Quantum effects in curved backgrounds and, in particular, the polarization of the quantum vacuum, may provide the necessary stress-energy tensor to support wormholes~\cite{Fabbri2005,Ho2017,Berthiere2017}.

It is generally accepted that standard particles of matter and waves can cross traversable wormholes without experiencing appreciable interactions with the exotic matter opening the throat, although there is virtually no knowledge about the possible interactions between standard matter and the source of the wormhole geometry. Hence, here we will consider that the interior of wormholes is essentially transparent, but keeping in mind that a deeper analysis of this issue would be desirable. This assumption would be certainly more reasonable if the exotic matter inside the wormhole comes entirely from the polarization of the quantum vacuum. This traversability property (or, in other words, the lack of a physical surface) represents the main difference between wormholes and quasi black holes as defined in the previous section. Aside from this difference, these two kinds of objects share several properties. In particular, most of (if not all) the uncertainties regarding the understanding of the dynamics of quasi black holes and their formation mechanisms equally apply to wormholes.

\section{Observational channels \label{sec:channels}}

\subsection{Phenomenological description of deviations from general relativity \label{sec:param}}

From the perspective of astrophysical observations using electromagnetic waves, black holes are regions in spacetime that can be detected only indirectly through their gravitational effects on matter surrounding them. This has changed with gravitational wave astronomy. However, observationally is not clear that these regions of spacetime correspond strictly to black holes in the sense of general relativity. This is a fundamental question regardless of the stance taken with respect to the different alternatives in Sec. \ref{sec:taxonomy}. Only a detailed analysis of this question would make possible separating what is really known from the aspects that can be only inferred from (most of the time, partial) theoretical arguments. 

In order to illustrate this point and make quantitative statements, let us introduce a set of phenomenological parameters encapsulating deviations from the behavior expected in general relativity. We will compare the physics associated with each of these parameters with the theoretical models reviewed in Sec. \ref{sec:taxonomy}, and then consider how these parameters can be constrained observationally. These parameters are functions of the physical quantities characterizing the most general black hole geometry that is expected to be relevant for astrophysical scenarios, namely the Kerr geometry~\cite{Kerr1963}: the mass $M$ and the angular momentum $J$ (that we do not deal explicitly with for simplicity). Let us start with two timescales:
\begin{enumerate}
\item{\emph{Lifetime}, $\tau_+$: The timescale in which a black hole with mass $M$, in vacuum, disappears completely (due either to Hawking radiation, or some other effect).}
\item{\emph{Relaxation}, $\tau_-$: is the amount of time in which $\mathscr{O}(1)$ transient effects taking place after violent dynamical processes dissipate (formation of the black hole, merger, etc.). Typically this can be identified with the imaginary part of the lowest quasinormal mode of the final-state system (e.g., \cite{Cardoso2003,Konoplya2011}).}
\end{enumerate}
These two timescales describe the interval of time $t\in[\tau_-,\tau_+]$ in which the system is expected to be evolving slowly enough that it can maintain stable structural properties. Within this time interval, it is meaningful to define the following parameters:
\begin{enumerate}\setcounter{enumi}{2}
\item{\emph{Size}, $R= r_{\rm s}(1+\Delta)$: Value of the radius below which the modifications to the classical geometry are $\mathscr{O}(1)$. We will use the more convenient parameter $\Delta\geq0$. Note that this parameter is related to $\mu$ in Eq.~\eqref{eq:mudef} as $\mu=\Delta/(1+\Delta)$. For $\Delta\ll1$ it follows that $\mu\simeq\Delta$, so that these two parameters can be used interchangeably.}
\item{\emph{Absorption coefficient}, $\kappa$: Measures the fraction of the energy that is semi-permanently lost inside the region $r\leq R$. This can be due to the inelastic interaction with the horizonless object, when exciting internal degrees of freedom in the bulk, or simply due to its propagation into some other spacetime region (consider, for instance, a wormhole).}
\item{\emph{Elastic reflection coefficient}, $\Gamma$: If there is a certain amount of energy falling onto the object and reaching $r=R$, this coefficient measures the portion that is reflected at $r\geq R$ due to elastic interactions (i.e., energy which is not absorbed and bounces back).}
\item{\emph{Inelastic reflection coefficient, $ \tilde{\Gamma} $}: Measures the portion of energy that is temporarily absorbed by the object and then re-emitted. That is, it measures the amount of energy that is inelastically reflected. It is related to $\kappa$ and $\Gamma$ by $\tilde{\Gamma}=1-\kappa-\Gamma$.}
\item{\emph{Tails}, $\epsilon(t,r)\ll1$: Modifications of the geometry that decay with radial distance, typically polynomial but possibly modulated by functions of compact support. Hence, there might be a maximum radius such that $\epsilon(t,r\geq r_\star)=0$. In principle one would need to introduce a series of functions $\epsilon^J(t,r)$ to describe different decaying tails for different coefficients of the metric~\cite{Johannsen2011}. {For $r_\star=\infty$ these tails would produce nonzero values of higher-order multipole moments (e.g., \cite{Cardoso2016b}).}}
\end{enumerate}
These phenomenological parameters (and functions) allow us to characterize our ignorance about the actual properties of astrophysical black holes: for a black hole in general relativity,
\begin{equation}\label{eq:param}
\begin{array}{lll}
\tau_+=\infty,&\qquad  \tau_-\sim 10 M, \qquad & \mu=0,\\
\kappa=1,&\qquad \Gamma=0,&\epsilon(t,r)=0.
\end{array}
\end{equation}
Regarding the first parameter $\tau_+$, we expect it to be at most $\tau_+=\tau^{(3)}$ as defined in Eq.~\eqref{eq:t3} due to Hawking radiation. But this is still infinite for any practical purposes for astrophysical black holes. The estimate for $\tau_-$ is obtained taking the inverse of the imaginary part of the lowest quasinormal mode which governs the damping rate of perturbations~\cite{Chandrasekhar1975,Kokkotas1999}. The rest of the parameters are unchanged in the semiclassical approximation. Hence testing the (semiclassical) black hole picture essentially means constraining the value of these parameters. The closer these parameters are to their values in Eq.~\eqref{eq:param}, the more confident we will be that astrophysical black holes are classical black holes (especially if we are able to discard some regions containing values associated with known theoretical models). The corresponding values for the various classes introduced in Sec. \ref{sec:taxonomy} are given in Table \ref{tab:par}. In the rest of the paper, we discuss the most stringent bounds that can be currently placed on these parameters.

It is always possible to introduce even more additional parameters or functional relations, such as frequency-dependent values of $\mu(\omega)$ and $\Gamma(\omega)$. However, in practical terms this just implies that we are including additional parameters that would provide more freedom to play with the observational data. The set we introduced above is minimal, but still interesting enough to give a detailed picture of the observational status of black holes. In practice, the only additional freedom that we will consider is the possibility that some of these coefficients are different for electromagnetic and gravitational waves.

\begin{center}
\begin{table*}[!htp]
\begin{center}
\begin{tabular}{||c||c|c||c||c|c|c||c||}
\hline
\hline
Model & $\tau_-$ & $\tau_+$ & $\Delta$ & $\kappa$ &$\Gamma$ & $\tilde \Gamma$ &$\epsilon(t,r)$ \\
\hline
\hline
{Classical black hole} & $\sim 10M$ & $\infty$ & 0 & 1 & 0 & 0 &0\\ 
\hline
Regular black hole & $\sim 10M$ & U & 0 & 1 & 0 &0 & MD \\ 
\hline
 Bouncing geometries & MD & MD & $0$  & 1 &0 &0 & $r_\star=\mathscr{O}(r_{\rm s})$ \\
\hline
 Quasi black hole & MD/U & $\infty$ & $>0$ & MD/U & MD/U & $1-\kappa-\Gamma$ & {U}  \\
\hline
 Wormhole & U & $\infty$ & $>0$ & MD & $1-\kappa$ &0& {U}  \\
\hline  
\hline  
\end{tabular}
\caption{Values of the phenomenological parameters for the different classes of black hole mimickers. MD stands for Model Dependent and U for Unknown, whereas MD/U emphasizes that the quantity is model dependent but at the moment there is no particular model within the class that is able to predict specific values for the corresponding parameters.}\label{tab:par}
\end{center}
\end{table*}
\end{center}
%

\subsection{Electromagnetic waves}

The presence of dark distributions of mass that do not themselves emit electromagnetic radiation can be indirectly detected by their gravitational effects on the surrounding luminous matter. This has been traditionally the strategy followed in order to hunt for black holes in astronomical data. This electromagnetic radiation can also be used to probe the gravitational field around these dark distributions of mass and even constrain some of their surface properties. In this section we review the most powerful observations and reevaluate their strength on the basis of the parametrization introduced in Sec.~\ref{sec:param} above.

\subsubsection{Orbiting stars \label{sec:orbits}}

The first kind of situation we shall consider can be idealized as a many-body system of compact distributions of matter that are interacting gravitationally, with at least one element that is not (appreciably) emitting electromagnetic radiation. The simplest possible configuration would be a binary system composed of a regular star and a dark companion, an example of which is A0620-00~\cite{Elvis1975} (which is also the closest system of this kind to the solar system). The electromagnetic radiation coming from the luminous star can be used in order to deduce the mass of its companion through the so-called mass function~\cite{McClintock1986,Casares2013}, which in the case of A0620-00 yields $6.60\pm 0.25\  M_\odot$~\cite{Cantrell2010}. This value is well above the maximum theoretically allowed mass for neutron stars~\cite{Oppenheimer1939,Rhoades1974,Chitre1976}. While the mass parameter can be extracted, all the phenomenological parameters introduced in Sec. \ref{sec:param} remain virtually unconstrained (or weakly constrained if compared with other observations detailed below). Hence, observations of these binary systems of stellar-mass objects justify the existence of dark and compact distributions of matter that are not neutron stars, but not much more information about the intrinsic properties of these structures can be extracted.

The situation may improve if the mass of the dark object increases by several orders of magnitude, which would typically imply moving from a binary system to a many-body system. The larger number of luminous stars improves the statistics and therefore allows placing stronger constraints. On the one hand, there is only a single system of this kind that is accessible to current technology: the center of our own galaxy. Sagittarius A* (Sgr A*) is an astronomical radio source at the center of the Milky Way, which has long been considered to be the location of an astrophysical black hole~\cite{Eckart2017}. On the other hand, this region has been extensively studied during more than two decades~\cite{Eckart1996,Ghez2000,Schodel2002,Ghez2003,Gillessen2008,Meyer2012}, with the result that the trajectories of a large number of orbiting stars are known with excellent precision~\cite{Boehle2016,Parsa2017}. A sizable portion of the claim that Sgr A* is a black hole comes in fact from these observations.

The main parameters which are fixed by these observations are the mass of Sgr A* and our distance from it. Precise values and errors can be found in the above references, but roughly these are given by $M=4\times 10^6\ M_\odot$ and $d=8\ \mbox{kpc}$, with errors of the order of 1\%. The measure of the distance obtained from tracking these stars, which is based on the geometric method proposed in~\cite{Salim1998}, is in good agreement with the results of other methods (e.g.,~\cite{Francis2014}). 

More interesting for the present discussion is the remark that these observations also constrain the size of Sgr A*, on the basis that these stars have been observed to travel freely without colliding with the central supermassive object (CMO in the following). The values of the periastron in these orbits provide upper bounds to the value of $R$ (equivalently, $\Delta$). For the purposes of estimating the order of magnitude of this quantity, it is enough to consider the star S2 (also known as S0-2)~\cite{Ghez2000,Schodel2002} which is one of the most precisely tracked. The periastron of S2 is 17 light hours, while the Schwarzschild radius of Sgr A* is 40 light seconds. Therefore,
\begin{equation}
\Delta\leq \mathscr{O}(10^3).\label{eq:orbstars1}
\end{equation}
This first bound is a very crude bound which will be tightened using other observational channels. Given that this kind of observation is essentially geometric in nature, the remaining phenomenological parameters that describe the physical intrinsic properties of the dark object remain unconstrained in practice. This continues to be true for more refined observations of the galactic center~\cite{Doeleman2008} that allow one to improve the constraint \eqref{eq:orbstars1} by about three orders of magnitude. In the future it may however be possible  to constrain the tails $\epsilon(t,r)$, though this would require gathering data for several stars with much shorter orbital periods~\cite{Will2007}, and which remain close enough to Sgr~A*~\cite{Merritt2009,Grould2017}. For instance, measurements of the redshift of S2, and their incompatibility with Newtonian mechanics, have been recently reported~\cite{Abuter2018}. However, distinguishing effects beyond general relativity would require much higher precision.

\subsubsection{Infalling matter close to the gravitational radius \label{sec:infalling}} 

Observations of stars orbiting CMOs at the center of galaxies are currently restricted to Sgr A* due purely to technological limitations, so that this constraint only applies to this particular astronomical source. Moreover, the distances involved in the orbits of the stars discussed in the previous section are large in comparison with the gravitational radius of the Sgr A*. However, another source of information comes from matter infalling on the CMO. It is reasonable to expect that processes involving matter in the surroundings of the gravitational radius constitute a better probe of the features of the CMO. 

In order to describe these processes, we need to briefly review some aspects of the behavior of geodesics around the gravitational radius of the CMO that are caused by the strong gravitational fields in the near-horizon region. Both ingoing and outgoing geodesics are interesting phenomenologically, as the former describe the approach of particles and waves to the CMO, while the latter describe how and when the radiation produced in different processes escapes from the gravitational field of the CMO. We can just focus on null geodesics, given that these determine the boundaries of the lightcones in which timelike geodesics have to be contained. As is usually done in spherical symmetry, we can restrict attention to the $\theta=\pi/2$ plane without any loss of generality, and reduce the geodesic equation for trajectories $x^\mu(\lambda)=(t(\lambda),r(\lambda),\pi/2,\varphi(\lambda))$ to
\begin{equation}\label{eq:nullg}
\left(\frac{\text{d}r}{\text{d}\lambda}\right)^2+\left(1-\frac{2M}{r}\right)\frac{L^2}{r^2}=E^2.
\end{equation}
The conserved quantities $E=(1-2M/r)\text{d}t/\text{d}\lambda$ and $L=r^2\text{d}\varphi/\text{d}\lambda$ correspond to the energy and angular momentum of the null geodesic. The derivation of these equations is described in most general relativity textbooks (see, for instance,~\cite{Carroll2004}). The second term in the left-hand side of the equation above acts as an effective potential. Circular trajectories ($\text{d}r/\text{d}\lambda=0$) can occur at maxima or minima of this effective potential, being respectively unstable or stable. It is straightforward to check from Eq.~\eqref{eq:nullg} that there is only one bound circular orbit, at
\begin{equation}
r_{\rm ph}=\frac{3}{2}r_{\rm s}=3M.
\end{equation}
The surface defined by $r=r_{\rm ph}$, known as the photon sphere, plays an important role in the discussions below.

Null geodesics that cross or reach the photon sphere have a maximum angular momentum $L_\star$ that can be directly evaluated from Eq.~\eqref{eq:nullg} by imposing the condition that $\left.(\text{d}r/\text{d}\lambda)^2\right|_{r=r_{\rm ph}}\geq0$,
\begin{equation}\label{eq:maxl}
L\leq L_\star=3\sqrt{3}ME.
\end{equation}
The main implication of the existence of this maximum angular momentum is that outgoing geodesics inside the photon sphere cannot cross the latter if $L>L_\star$. A similar comment applies to ingoing geodesics outside the photon sphere.

\begin{figure}[t]%
\begin{center}
\vbox{\includegraphics[width=0.3\textwidth]{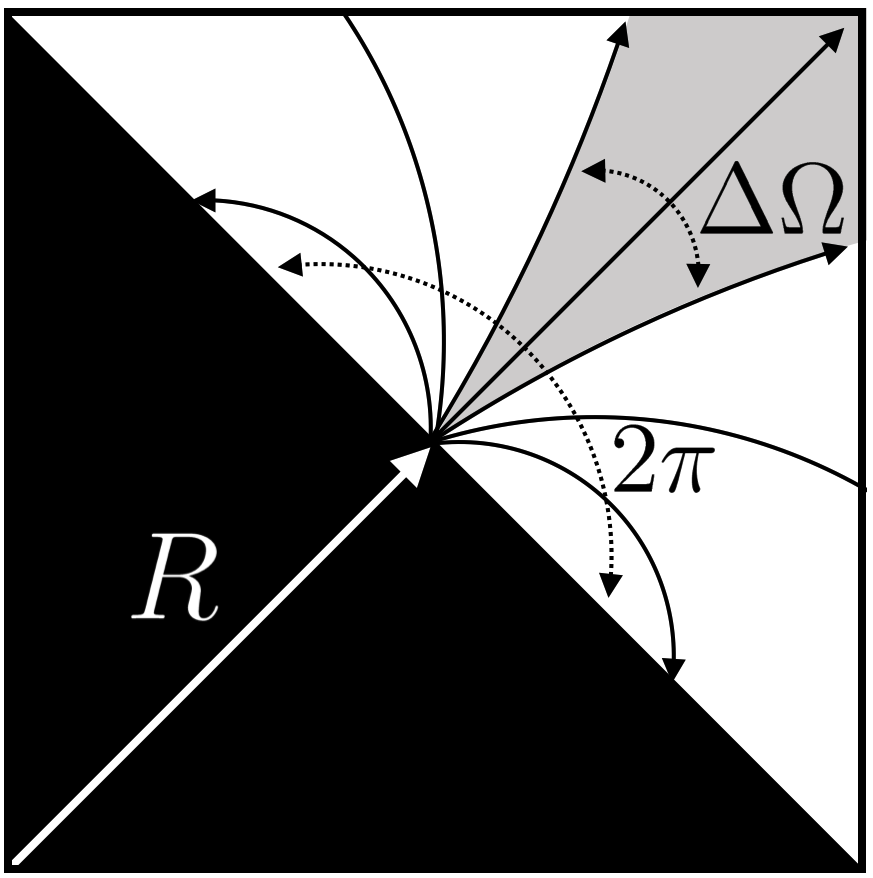}}
\bigskip%
\caption{Only a fraction $\Delta\Omega/2\pi$ of geodesics emitted isotropically at a point on the surface $r=R$ can escape for ultra-compact configurations.}
\label{fig:lensing}%
\end{center}
\end{figure}%

Let us now consider for instance an object with a surface at $r=R\leq r_{\rm ph}$ such that every point on the surface emits electromagnetic radiation isotropically in its local orthonormal frame $\{e^\mu_t,e^\mu_r,e^\mu_\theta,e^\mu_\varphi\}$. A fraction of these initially outgoing rays cannot reach the photon sphere, which means (see Fig. \ref{fig:lensing}) that these will be strongly curved and will come back to the surface $r=R$~\cite{Abramowicz2002} (see also~\cite{Cardoso2017}). The escape angle $\vartheta_\star$ measured from the normal to the surface can be determined imposing the critical value $L=L_\star$ and calculating
\begin{align}
\sin\vartheta_\star&=\left.\frac{g_{\mu\nu}e^\mu_\varphi \text{d}x^\nu/\text{d}\lambda}{\sqrt{\displaystyle\left(g_{\mu\nu}e^\mu_\varphi \frac{\text{d}x^\nu}{\text{d}\lambda}\right)^2+\left(g_{\mu\nu}e^\mu_r \frac{\text{d}x^\nu}{\text{d}\lambda}\right)^2}}\right|_{r=R,\,\theta=\pi/2,\,L=L_\star}
\\
&=\frac{L_\star}{ER}\sqrt{1-\frac{2M}{R}}.
\end{align}
Here we have used $e^\mu_r=(0,\sqrt{1-2M/r},0,0)$, and $e^\mu_\varphi=(0,0,0,1/r)$, and we keep $\theta=\pi/2$ without loss of generality. The solid angle spanned by the cone of geodesics that escape from the sphere $r=R$ can be then calculated as 

\begin{align}
\Delta\Omega&=\int_0^{2\pi}\text{d}\varphi\int_0^{\vartheta_\star}\text{d}\vartheta\sin\vartheta=2\pi(1-\cos\vartheta_\star)\\
&=2\pi\left[1+ \left(1-\frac{3M}{R}\right) \sqrt{1+\frac{6M}{R}}\,\right].
\label{eq:lensang}
\end{align}
In the limit $R\rightarrow r_{\rm s}=2M$ (in which $\Delta\simeq\mu\ll1$), one has
\begin{equation}\label{eq:sol_ang}
\frac{\Delta\Omega}{2\pi}=\frac{27}{8}\mu+\mathscr{O}(\mu^2).
\end{equation}
Therefore, only a small fraction of the light emitted from the surface of the object will escape to infinity for ultracompact configurations. After this important remark, we can study two different ways in which matter falls onto the CMO, namely the case in which stars collide with the CMO (triggering a ``stellar disruption event''~\cite{Lu2017}) and the case in which the CMO is surrounded by an accretion disk.

\paragraph{\textbf{Stellar disruption events:} \label{sec:sde}}

The physics associated with the possible collision of an orbiting star with the CMO gets quite complicated due to the existence of tidal forces. For a given pair, of star and CMO, there is a critical value of the radius $r_{\rm T}$ known as the Roche limit (or tidal radius), in which the internal forces holding the star together cannot endure the gravitational tidal forces, and the star is torn apart. A Newtonian estimate of the order of magnitude of this radius is $r_{\rm T}\sim R_\star(M/M_\star)^{1/3}$, where $R_\star$ and $M_\star$ are the radius and the mass of the orbiting star, and $M$ is the mass of the CMO~\cite{Rees1988}. For $M\gtrsim 10^7 M_{\odot}$ the tidal radius is very close to the Schwarzschild radius, therefore, tidal disruption events (TDE) happen in a region in which the Newtonian treatment is not sufficient and relativistic tidal forces must be taken into account~\cite{Kesden2011}, which is associated with the relativistic nature of the near-horizon orbits. Moreover, for $M\gtrsim 10^8 M_{\odot}$ (this value for the order of magnitude takes into account the relevant relativistic features), tidal forces are not strong enough, so that main-sequence stars are able to reach the Schwarzschild radius while keeping their integrity~\cite{Servin2016,MacLeod}.

That tidal disruption events (TDEs) have been observed for CMOs of $M\sim 10^6M_\odot$~\cite{Komossa2015} leads to a first upper bound $r_{\rm s}(1+\Delta)\leq r_{\rm T}$ for these CMOs~\cite{Lu2017}. Taking as a reference $M_\star\sim M_\odot$ and $R_\star\sim R_\odot$, one obtains 
\begin{equation}
\Delta\leq\mathscr{O}(10).\label{eq:tde1}
\end{equation}
This improves by two order of magnitudes the bound \eqref{eq:orbstars1} that applies to the same value of the mass, $M\sim 10^6 M_\odot$. But it is possible to do even better if we move to the range of masses between $M\sim 10^8 M_\odot$ and $M\sim 10^{10} M_\odot$, for which there are no TDEs and the descending star is allowed to continue its trip downwards and reach the radius $R=r_{\rm s}(1+\Delta)$. The star would then crash into the surface, producing an envelope of debris that will radiate its energy away at the Eddington luminosity. The corresponding temperature at infinity is given by
\begin{equation}
T_\infty=\left(\frac{L_{\rm Edd}}{4\pi\sigma_{\rm SB} R^2}\right)^{1/4}\left(\frac{\Delta\Omega}{2\pi}\right)^{1/4}.\label{eq:tlum}
\end{equation}
This is just the Stefan-Boltzmann law applied to the Eddington luminosity $L_{\rm Edd}$  integrated over the area of a sphere with radius $R$, and suppressed by the geometrical fraction $\Delta\Omega/2\pi$ of radiation that actually escapes to spatial infinity. The value of Eddington luminosity depends on the properties of the accreting matter but, in situations in which the two relevant parameters are the molecular mass $m$ of the gas and the scattering cross-section $\sigma$ between photons and gas particles, dimensional arguments lead to $L_{\rm Edd} \propto G M m c/\sigma$, where the proportionality factor should be taken to be $4\pi$ in order to obtain the usual result \cite{Padmanabhan2001}. The Stefan-Boltzmann constant is $\sigma_{\rm SB}\simeq5.67\times10^{-8}\mbox{W m}^{-2}\mbox{ K}^{-4}$. We refer to~\cite{Lu2017} for a detailed discussion of these aspects, while focusing the present discussion on the universal dependence of the equation above on the geometrical factor $\Delta\Omega/2\pi$.

If we combine Eqs. \eqref{eq:sol_ang} and \eqref{eq:tlum}, we see that the temperature of the envelope of debris goes to zero as $\mu^{1/4}$. This makes harder to probe this phenomenon the more compact the CMO is. This feature is characteristic of inelastic processes in which some energy interacts with the surface of the CMO and is then radiated isotropically in the corresponding local reference frame, hence suffering the lensing effects described in Sec. \ref{sec:infalling}.

This luminosity can be constrained using astronomical surveys, in particular the Pan-STARRS1 3$\pi$ survey~\cite{Magnier2013}. The larger the value of $\mu$, the larger the luminosity, so that this analysis should lead to an upper bound on the value of $\mu$. In order to do so, one needs additional information about the number of CMOs with a given mass and for a given value of redshift, and also an estimation of the number of stellar disruption events that would occur. The details of the distribution of the layer of debris around the CMO and, in particular, the position of the photosphere of this envelope, are also important. Taking into account all these details, the authors of~\cite{Lu2017} obtain a constraint $\mu\leq \mu_{\rm CMO}=10^{-4}$ (we have just rounded off the value of the exponent). Note that the electromagnetic radiation is emitted from the photosphere, so that this observational channel is ultimately placing constraints on the size of the latter. We can remove the effect of the complex details regarding the thickness of the layer of debris by considering instead a very conservative bound derived from the fact that the radius of the CMO must certainly be smaller than the radius of the photosphere, namely $\mu\leq \mu_{\rm ph}$ where $\mu_{\rm ph}$ measures the size of the photosphere. In order to do so, we just need to take into account that $\mu_{\rm ph}/\mu_{\rm CMO}\simeq \kappa_{\rm T} M_\star/4\pi r_{\rm s}^2$ (see~\cite{Lu2017} for the derivation), with $\kappa_{\rm T}=0.34\times 10^{-3}\mbox{ m}^2\mbox{ kg}^{-1}$ the Thomson opacity for solar metallicity, and $M_\star=\mathscr{O}(M_\odot)$. We can then write
\begin{equation}
\mu\leq 10^{-4}\frac{\kappa_{\rm T} M_\star}{4\pi r_{\rm s}^2}= \mathscr{O}(1)\times\left(\frac{10^8 M_\odot}{M}\right)^2.\label{eq:sdecons}
\end{equation}
It is worth stressing that this bound still relies on a series of significant assumptions regarding the cosmological population of CMOs and the rate of stellar disruption events (and also an assumption that $\Delta$ does not depend explicitly on the mass of the CMO, an assumption which should be relaxed in future analyses). It should be therefore taken as a first estimate, and as a proof of principle that this kind of observation can be used to constrain the phenomenological parameters discussed here. More refined analysis and future observations would help to strengthen the accuracy of this bound. 

Most importantly, Eq.~\eqref{eq:sdecons} assumes that $\kappa$ and $\Gamma$ are both vanishing. The introduction of nonzero values for these two parameters has a significant impact on the discussion, with the left-hand side of Eq.~\eqref{eq:sdecons} picking up factors that depend explicitly on these phenomenological parameters. The change in this equation is functionally equivalent to the change of the upper bound discussed in the next section, with the general outcome that the upper bound on $\mu$ becomes weaker for nonzero values of these parameters. We will show this explicitly in the discussion below, that includes naturally all the steps that are needed in order to take these parameters into account.

One last comment is that the factor that depends on $\Delta\Omega/2\pi$ in Eq.~\eqref{eq:tlum} is essential in order to avoid running into significantly problematic and wrong conclusions. Ignoring this factor and writing $T_\infty=(L_{\rm Edd}/4\pi\sigma_{\rm SB} R^2)^{1/4}$ would instead have resulted in an overestimate of the the outgoing flux of radiation by several orders of magnitude. It is clear that this would had led to stronger (but nevertheless flawed) constraints than Eq.~\eqref{eq:sdecons}.

\paragraph{\textbf{Accretion disks around supermassive black holes:} \label{sec:supacc}}

The most stringent constraints on some of the phenomenological parameters come from the information about the average amount of infalling matter per unit of time onto CMOs. The value of this accretion rate $\dot{M}$ is generally more stable than the (much higher, but also more variable) accretion rate associated to the (much rarer) direct capture of an individual star. Estimations of the accretion rate for these objects depend on the physics of accretion disks~\cite{Quataert1999,Hawley2002,Narayan2005}, as the accretion rate is typically estimated from the luminosity of the disk. As we have done in the previous section, we will not discuss the model-dependent features behind these estimations. We will just assume that it is possible to obtain a measure of the order of magnitude of $\dot{M}$, focusing our discussion on the (already rich) physics that can be described in terms of $\dot{M}$ and our phenomenological parameters introduced previously. More accurate estimations of $\dot{M}$ would just permit to refine the observational bounds given below.

Let us start summarizing the main argument that has been invoked several times in the literature~\cite{Broderick2005,Broderick2007,Broderick2009,Narayan2008,Broderick2015}. We can reduce this argument to its essentials by considering the system composed by the CMO and the accretion disk as a composite system in which energy is exchanged between its two components. The accretion rate $\dot{M}$ measures the energy that the accretion disk is pumping into the CMO. On the other hand, the quantity that is interesting in order to test the nature of the CMO is the energy that the CMO emits by itself, as this measures the reaction of the CMO to its interaction with the accretion disk. Ideally, one would like to disentangle the two fluxes of energy and measure independently the radiation emitted by the CMO. However, this is not yet observationally possible (and, as we discuss below, might be even impossible in practice due to its extreme faintness). Therefore, it is necessary to make additional assumptions in order to determine the properties of this outgoing energy flux:
\begin{enumerate}
\item{\emph{Thermality:} It was pointed out in~\cite{Broderick2005,Broderick2007,Broderick2009} that the strong lensing of outgoing geodesics emitted at different points in the surface $r=R$  (a phenomenon that we have discussed in Sec. \ref{sec:infalling}) implies that the surface reaches thermal equilibrium on a short timescale. This follows from the fact that different points of the surface are strongly coupled. Therefore, we can safely assume that the emitted radiation is thermal. This is correct for $\Delta\simeq\mu\ll1$, which means that the arguments in this section hold only in these situations.}
\item{\emph{Steady state:} The only parameter to be fixed after accepting the assumption above is the temperature $T_\infty$ of the emitted radiation. The only possible model-independent argument to fix the power of the outgoing radiation is invoking conservation of energy and assuming that a steady state between the two components (CMO and accretion disk) has been reached, so that the two fluxes of energy carry the same amount of energy. This is a strong assumption that must be appropriately justified and that, as we see below, \emph{fails} due to several competing effects.}
\end{enumerate}

If these two assumptions hold, then the emission of the CMO can be calculated: It is given by a thermal distribution with a temperature determined by the accretion rate $\dot{M}$. For Sgr A*, this radiation should be bright in the infrared, but it has been shown~\cite{Broderick2005,Broderick2007,Broderick2009,Narayan2008} that the emission of Sgr A* in the infrared is about $10^{-2}$ times this theoretical estimate (see~\cite{Broderick2015} for the same argument applied to M87). The conclusions by these authors is that it is not possible that Sgr A* has a surface, and therefore that it must have a horizon. This is a very strong claim, as it would discard every possible value of $\Delta$, except for $\Delta=0$ which corresponds to a black hole. Even sub-Planckian values of $\Delta$ would be discarded. Let us now analyze in careful detail how this conclusion comes about --- so to dispel  any possible doubt concerning its robustness.

An obvious starting point for this revision are the two assumptions mentioned above. Indeed, as we show below, the Achilles' heel of this argument is the steady state assumption. This assumption is not valid for sufficiently compact CMOs; which leads to constraints on the maximum compactness of the object. This observation has been made only recently~\cite{Lu2017,Cardoso2017nat}. It is worth mentioning that even recent reviews on this topic such as~\cite{Eckart2017} still quote~\cite{Broderick2005,Broderick2007,Broderick2009,Narayan2008,Broderick2015} as definitive evidence for the existence of event horizons, thus ignoring this loophole.\footnote{It is perhaps worth stressing again that no local observation in space and time will be ever able to observe an event horizon, which is intrinsically a teleological notion. At most, observations will be able to confirm or exclude the existence of trapping horizons or other local definitions of the boundary of black holes. See, e.g.,~\cite{Visser2014} for an extensive discussion of this point.} Our novel contribution to this discussion is the introduction (in terms of the phenomenological parameters defined in Sec. \ref{sec:param}) of additional physical features that are expected to be relevant in realistic scenarios. As we discuss below, the introduction of these additional physical aspects makes these constraints significantly weaker. 

\begin{figure}[h]%
\begin{center}
\vbox{\includegraphics[width=0.5\textwidth]{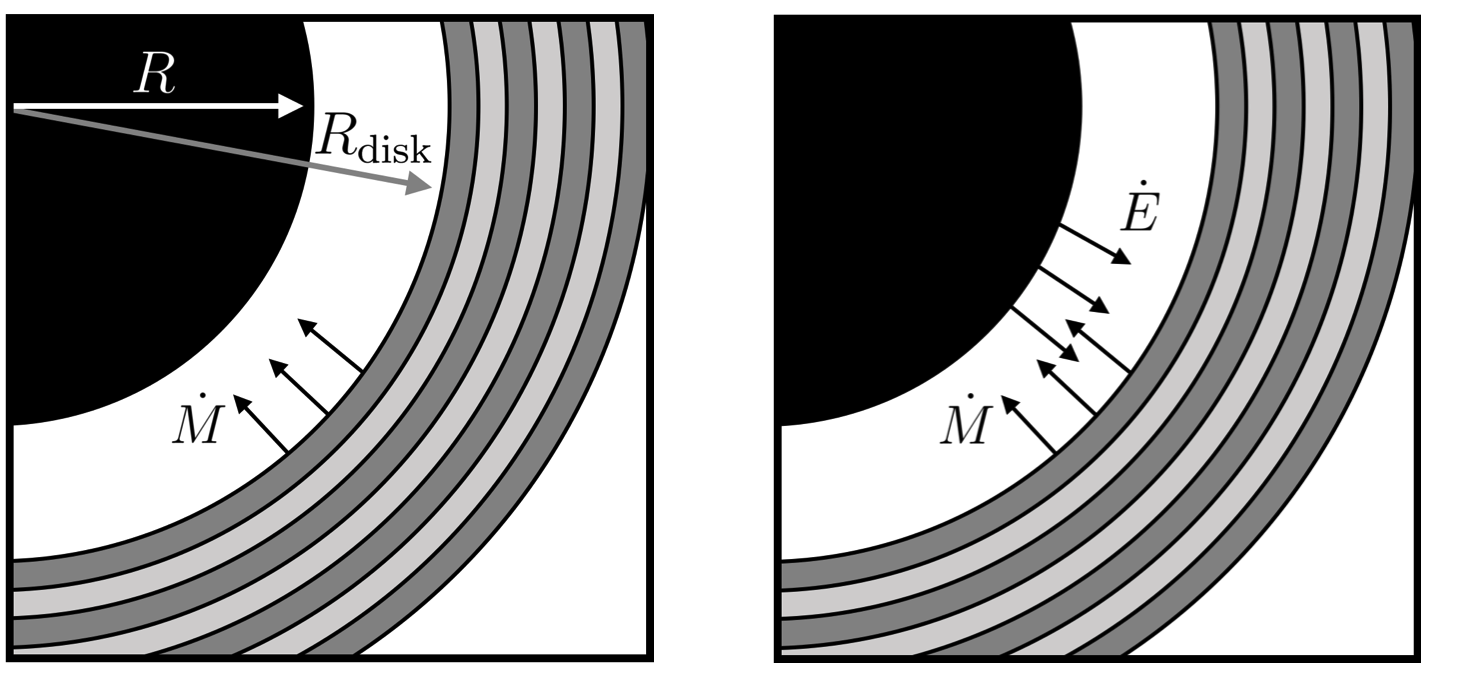}}
\bigskip%
\caption{On the left: Initial state in which matter starts falling at a rate $\dot{M}$ from the accretion disk onto the CMO. On the right: steady state in which the energy emitted from the CMO and reaching the accretion disk is $\dot{E}=\dot{M}$.}
\label{fig:twosteady}%
\end{center}
\end{figure}

Let us consider a simple calculation of the time at which steady state is reached (see Fig.~\ref{fig:twosteady}). The initial configuration is given by an accretion disk that starts pumping energy into the CMO,  the energy emission of the latter being negligible before accretion begins. We will start considering the most favorable case in which the CMO returns all the accreted energy as thermal radiation, and evaluate the timescale at which steady state can be achieved. Hence, the accretion rate onto the CMO is zero for $t<0$ (this is just an irrelevant choice of the origin of time) and $\dot{M}\in\mathbb{R}$ for $t\in[0,T]$, where the timescale $T$ is short enough so that the approximation of constant $\dot{M}$ is reasonable (more details below). For simplicity, we assume that all propagating energy is carried along null geodesics, and also restrict the discussion to spherically symmetric situations. The amount of energy emitted per unit time by the CMO, $\dot{E}$, is measured at the location of the accretion disk $r=R_{\rm disk}$. Our goal is describing its evolution for $t\geq0$. There are two effects to keep into account. First of all, the energy emitted $\dot{E}$ remains negligible until the first ingoing radial null geodesics can bounce back at the surface $r=R$ and return to the accretion disk. This time can be directly evaluated using the Schwarzschild metric as
\begin{equation}
T_{\rm bounce}=2\left[R_{\rm disk}-R+r_{\rm s}\ln\left(\frac{R_{\rm disk}-r_{\rm s}}{R-r_{\rm s}}\right)\right].\label{eq:tbounce}
\end{equation}
This timescale is divergent in the limit $R\rightarrow r_{\rm s}$, or equivalently $\Delta\rightarrow0$. However, the logarithmic behavior implies that even for extremely small, but strictly non-vanishing $\Delta$, Eq.~\eqref{eq:tbounce} would be at most $\mathscr{O}(10)\times r_{\rm s}$. Hence, this timescale is essentially the light-crossing time of the CMO.

This effect alone would delay the moment in which the steady state would be reached, but, given the logarithmic dependence, even sub-Planckian values for $R-r_{\rm s}$ would be ruled out. However, there is a second effect to keep into account. Outgoing null geodesics are strongly lensed, which implies that a fraction of them do not escape and fall again onto the surface of the CMO. This effect is unavoidable due to the inherently inelastic nature of the process that is necessary for thermalization to take place: the energy falling from the accretion disk is absorbed by the CMO in the first place, and then emitted. Even assuming spherical symmetry for the infalling energy, particles would not hit the surface and bounce back radially. On the contrary, this emission would be isotropic in a local frame at rest in the surface, thus implying that only a very small fraction of the initially absorbed energy contributes to $\dot{E}$. The remaining energy follows highly curved trajectories and is reabsorbed by the CMO in a timescale that can be calculated numerically and is also controlled by its Schwarzschild radius, being $\mathscr{O}(10)\times r_{\rm s}$ at most. Then, a repetition of this process takes place, until eventually all the energy is radiated away.

\begin{figure}[h]%
\begin{center}
\vbox{\includegraphics[width=0.5\textwidth]{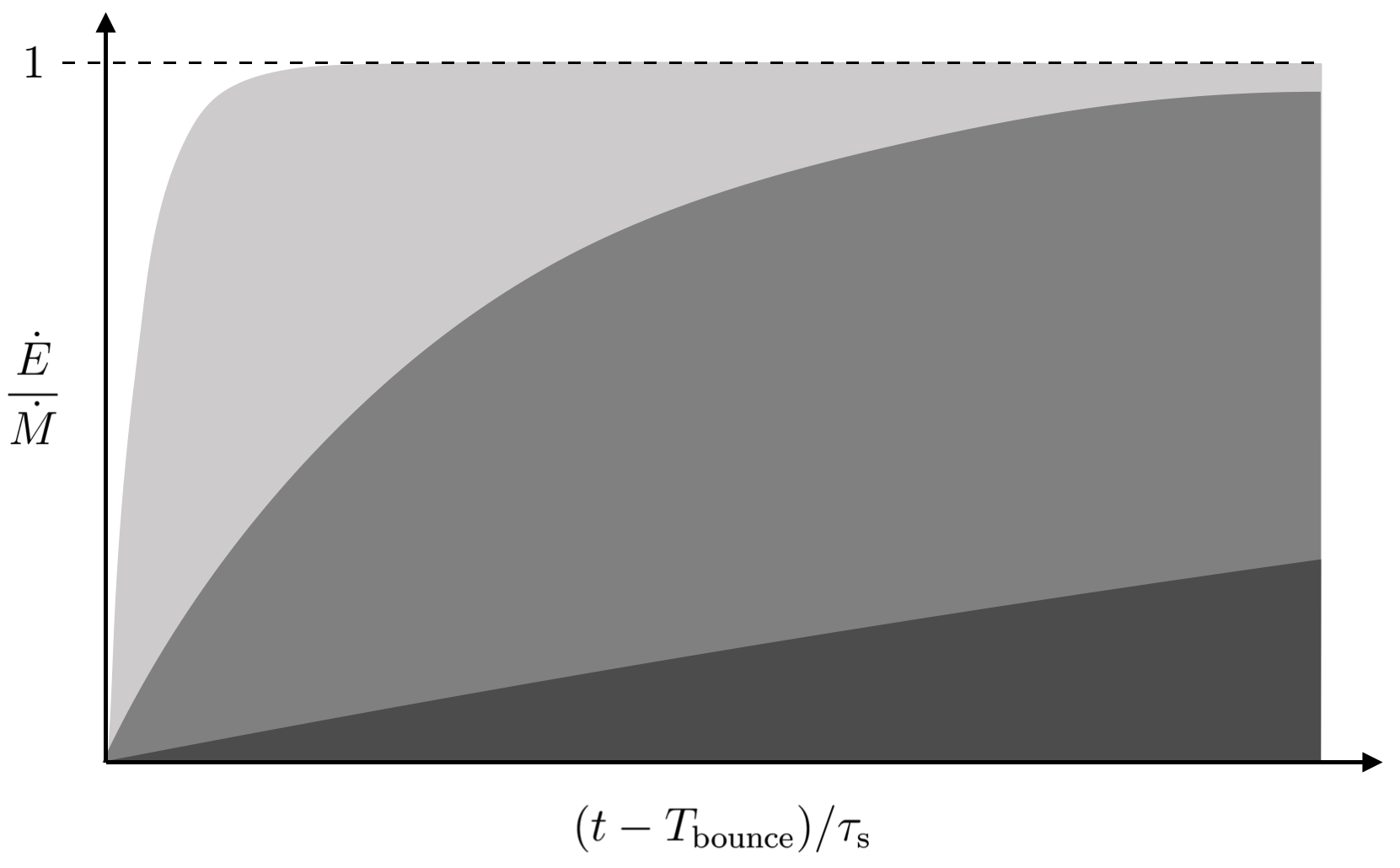}}
\bigskip%
\caption{Representation of Eq.~\eqref{eq:enrate} for $\Delta=0.1$ (light gray), $\Delta=0.01$ (gray) and $\Delta=0.001$ (dark gray).}
\label{fig:steady}
\end{center}
\end{figure}

In order to make the calculation tractable, let us follow the discussion in~\cite{Cardoso2017} and consider discrete intervals with their size given by the characteristic timescale $\tau_{\rm s}=\mathscr{O}(10)\times r_{\rm s}$, starting at $t=T_{\rm bounce}$. During each of these intervals, the mass that the accretion disk is ejecting into the CMO is given by $\dot{M}\tau_{\rm s}$. In the first interval after $T_{\rm bounce}$, the amount of outgoing energy that reaches the accretion disk is given by the corresponding fraction of the first injection of energy,
\begin{equation}\label{eq:seed1}
E_1=\frac{\Delta\Omega}{2\pi}\dot{M}\tau_{\rm s}.
\end{equation}
During the second interval, one would get the same fraction of the energy corresponding to the second injection, plus a fraction of the remaining energy from the first injection:
\begin{align}
E_2&=\left[\frac{\Delta\Omega}{2\pi}+\frac{\Delta\Omega}{2\pi}\left(1-\frac{\Delta\Omega}{2\pi}\right)\right]\dot{M}\tau_{\rm s}
\nonumber\\
&=  E_1 + \left(1-\frac{\Delta\Omega}{2\pi}\right) E_1.
\end{align}
In general, one can show that
\begin{equation}\label{eq:ener0}
E_n=\sum_{k=1}^n\epsilon_k,
\end{equation}
where the partial energies can be determined from the recurrence relation
\begin{equation}\label{eq:rec1}
\epsilon_{k+1}=\left(1-\frac{\Delta\Omega}{2\pi}\right)\epsilon_k,\qquad k\geq 1,
\end{equation}
with the seed $\epsilon_1=E_1$ given in Eq.~\eqref{eq:seed1}. Summing the geometric series, it follows then that
\begin{equation}\label{eq:ener}
E_n=\frac{\Delta\Omega}{2\pi}\dot{M}\tau_{\rm s}\sum_{k=0}^{n-1}\left(1-\frac{\Delta\Omega}{2\pi}\right)^k=\dot{M}\tau_{\rm s}\left[1-\left(1-\frac{\Delta\Omega}{2\pi}\right)^n\right].
\end{equation}
The accretion rate $\dot{M}$ is obtained dividing the mass accreted in each of these intervals by $\tau_{\rm s}$. Therefore, let us analogously define $\dot{E}_n=E_n/\tau_{\rm s}$.
When $\tau_s\ll T$, the timescale during which the accretion rate $\dot{M}$ is roughly constant, we can formally take the limit in which the size of the time intervals goes to zero and therefore $\dot{E}_n$ becomes a function of a continuous variable, $\dot{E}(t)$, which can be written in terms of the continuous variable $t\in[T_{\rm bounce},T]$ as
\begin{equation}
\frac{\dot{E}(t)}{\dot{M}}=1-\left(1-\frac{\Delta\Omega}{2\pi}\right)^{(t-T_{\rm bounce})/\tau_{\rm s}}.\label{eq:enrate}
\end{equation}
There are certain limits that are illustrative of the physics behind Eq.~\eqref{eq:enrate} (see also Fig. \ref{fig:steady}):
\begin{itemize}
\item{In the limit $R\rightarrow r_{\rm ph}=3r_{\rm s}/2$, one has $\Delta\Omega/2\pi\rightarrow1$ [recall Eq.~\eqref{eq:lensang}]. This implies that $\dot{E}=\dot{M}$ identically for $R\geq r_{\rm ph}$. In this limit, relativistic lensing effects disappear: for a regular star (neutron star or less dense), if the surface of the star emits instantly the absorbed energy, then after a large enough timescale (with respect to $T_{\rm bounce}$) the system reaches a steady state. It was this very same intuition originated in these astrophysical systems that led to the authors of the works~\cite{Broderick2005,Broderick2007,Broderick2009} to assume that the steady state is reached in this same timescale for CMOs of arbitrary compactness.}
\item{In the limit $R\rightarrow r_{\rm s}$ ($\Delta\rightarrow 0$) one has \mbox{$\dot{E}/\dot{M}\rightarrow 0$}. This corresponds to the known astrophysical behavior of a black hole, in which a steady state cannot be achieved~\cite{Broderick2005,Broderick2007,Broderick2009}. However, this limit is not abrupt, but proceeds in a continuous way: for $\Delta\simeq\mu\ll1$, one has
\begin{equation}
\frac{\dot{E}}{\dot{M}}\simeq\mu(t-T_{\rm bounce})/\tau_{\rm s}.
\end{equation}
In particular, there is a maximum value of $\dot{E}$ that is determined from the equation above when $t=T$ (if the accretion rate changes, the system would have to adapt to the new accretion rate and therefore the process of stabilization would restart).
}
\end{itemize}
%

The second limit above illustrates that relativistic lensing effects cannot be ignored for $\mu\ll1$, and can indeed spoil the stabilization of the composite system into a steady state. In particular, for Sgr A* the typical timescale for the variation of its accretion rate is set by the Eddington timescale $T=Mc^2/L_{\rm Edd}\simeq 3.8\times10^8\mbox{ yr}$. Hence, given that the emission of Sgr A* is at most $10^{-2}$ times that predicted under the steady state assumption~\cite{Broderick2005}, we can write
\begin{equation}
\left.\frac{\dot{E}}{\dot{M}}\right|_{t=T}\simeq\mu(T-T_{\rm bounce})/\tau_{\rm s} \leq\mathscr{O}(10^{-2}).\label{eq:acccons0}
\end{equation}
Plugging the numbers into this equation, we obtain
\begin{equation}
\mu\simeq \Delta \leq\mathscr{O}(10^{-17}).\label{eq:acccons}
\end{equation}
In particular, we see that the steady state assumption is not valid if $\mu$ satisfies this constraint. In other words, this constraint would be the strongest statement that can be made using this method.

It is interesting to translate this constraint into length scales. It implies that it is possible to rule out the existence of a surface that emits all the absorbed energy as thermal radiation with a precision of $10^{-17}$ (in the coordinate distance $r$) on the size of the CMO. In terms of proper radial distances, this precision becomes smaller due to the Schwarzschild factor $\mu=1-r_{\rm s}/R$, and is in fact roughly of $10^2$ meters over a size of $10^{10}$ m, which is certainly impressive. On the other hand, this is still more than 70 orders of magnitude greater than $\Delta\sim \ell_{\rm P}^2/r_{\rm s}$ (corresponding to a proper radial distance of the order of the Planck length).

The same argument (although without taking into account the lensing of the geodesics in the near-horizon region) has been applied to the CMO in M87~\cite{Broderick2015}, which is three orders of magnitude more massive than Sgr A*~\cite{Oldham2016}. Taking into account the adjustments discussed in this section, we can find a constraint that is several orders of magnitude weaker than the one that applies to Sgr A*.

Most importantly, it is natural to expect that the surface of the CMO will not strictly have $\kappa=\Gamma=0$. As we now show, the introduction of these parameters describing additional physics regarding the nature of the CMO has a large impact in the discussion, with $\kappa$ having the largest impact. 

Intuitively, the reason for this is clear. Before escaping the gravitational field of the CMO, radiation undergoes several cycles of absorption (after being lensed back to the CMO) and emission. If $\kappa\neq 0 $, in each of these cycles only a fraction $(1-\kappa)$ of the absorbed energy is emitted, which suppresses the overall power of the radiation emitted by the CMO. Let us write explicitly the main equations for $\kappa\neq0$. Eq.~\eqref{eq:ener0} still holds, but the recurrence relation \eqref{eq:rec1} is modified to
\begin{equation}
\epsilon_{k+1}=(1-\kappa)\left(1-\frac{\Delta\Omega}{2\pi}\right)\epsilon_k,\qquad k\geq1,
\end{equation}
now with seed $\epsilon_1=(1-\kappa)\Delta\Omega/2\pi\dot{M}r_{\rm s}$. It follows that
\begin{widetext}
\begin{equation}\label{eq:enratek}
\frac{\dot{E}}{\dot{M}}=\frac{(1-\kappa)\Delta\Omega/2\pi}{\kappa+(1-\kappa)\Delta\Omega/2\pi}\left[1-\left(1-\kappa\right)^{(t-T_{\rm bounce})/\tau_{\rm s}}\left(1-\frac{\Delta\Omega}{2\pi}\right)^{(t-T_{\rm bounce})/\tau_{\rm s}}\right].
\end{equation}
\end{widetext}
We see that the power $\dot{E}$ is nonlinearly suppressed with $\kappa$. In the limit in which $t\rightarrow\infty$, that is $t/\tau_s\gg\kappa^{-1}$, we get
\begin{equation}\label{eq:limtinf}
\lim_{t\rightarrow\infty}\frac{\dot{E}}{\dot{M}}=\frac{(1-\kappa)\Delta\Omega/2\pi}{\kappa+(1-\kappa)\Delta\Omega/2\pi}.
\end{equation}
For $\kappa=0$ identically, the equation above becomes \mbox{$\dot{E}=\dot{M}$}, which means that steady state is certainly reached if waiting for infinite time. However, for values of $\kappa$ that are still small but satisfy $\Delta\Omega/2\pi\ll\kappa\ll1$ (note that $\Delta\Omega/2\pi\ll1$ in order to guarantee that the emitted radiation is thermal), we see that
\begin{equation}
\lim_{t\rightarrow\infty}\frac{\dot{E}}{\dot{M}}\simeq \frac{\Delta\Omega/2\pi}{\kappa}\ll 1.
\end{equation}
In other words, the transfer of energy from surface degrees of freedom to bulk degrees of freedom strongly dampens the thermal emission from the surface of the CMO. Instead of Eq.~\eqref{eq:acccons}, we obtain then the much weaker constraint
\begin{equation}\label{eq:acccons2}
\frac{\mu}{\kappa}\leq \mathscr{O}(10^{-2}).
\end{equation}
This equation can be understood as a lower bound on the value of $\kappa$ that makes hard surfaces that would otherwise be excluded by  Eq.~\eqref{eq:acccons} compatible with the available observational data. 

Let us consider for instance the value  $\mu=\mathscr{O}(10^{-7})$ that is 10 orders of magnitude greater than the constraint \eqref{eq:acccons}, which is valid only for $\kappa=0$. From Eq.~\eqref{eq:acccons2}, we see that an absorption coefficient as small as
\begin{equation}
\kappa\geq \mathscr{O}(10^{-5})
\end{equation}
makes the existence of such surfaces compatible with observations.

It is also possible to obtain the equivalent of Eq.~\eqref{eq:enratek} for $\Gamma\neq0$. The only difference is that the recurrence relation is in this case
\begin{equation}
\epsilon_{k+1}=\left[(1-\kappa)\left(1-\frac{\Delta\Omega}{2\pi}\right)+\Gamma\frac{\Delta\Omega}{2\pi}\right]\epsilon_k,\qquad k\geq 2,
\end{equation}
and the seed of this relation is modified to
\begin{equation}
\epsilon_1=(1-\kappa-\Gamma)\frac{\Delta\Omega}{2\pi}\dot{M}r_{\rm s},\qquad \epsilon_2=(1-\kappa-\Gamma)\left(1-\frac{\Delta\Omega}{2\pi}\right)\epsilon_1.
\end{equation}
From these equations it is possible to check that a nonzero value of $\Gamma$ further weakens these constraints, although this effect is not as pronounced as the one associated with the absorption coefficient $\kappa$ given that it will not produce an exponential suppression like the one in Eq. \ref{eq:enratek}. For completeness, let us note that the analogue of Eq. \eqref{eq:limtinf} can be shown to be
\begin{equation}
\lim_{t\rightarrow\infty}\frac{\dot{E}}{\dot{M}}=\frac{(1-\kappa-\Gamma)(1-\Gamma)\Delta\Omega/2\pi}{\kappa+(1-\kappa-\Gamma)\Delta\Omega/2\pi}.
\end{equation}
We see that wormholes represent an extreme case from this perspective, as $\kappa+\Gamma=1$ and therefore $\dot{E}=0$ identically. Hence, wormholes cannot be tested as black hole alternatives using this particular observation channel.

\paragraph{\textbf{Consistency constraints from accretion:} \label{sec:conscon}}

Even in the best-case (but unphysical) scenario in which $\kappa=\Gamma=0$, Eq.~\eqref{eq:acccons} should be improved by about 70 orders of magnitude in order to rule out well-motivated theoretical values of $\mu$ such as the one that follows from $\Delta\sim \ell_{\rm P}^2/r_{\rm s}$ and that can be obtained from Eq. \eqref{eq:muquad}. Such an improvement of observational data seems hardly realistic, thus suggesting that certain theoretical models are almost impossible to probe. The situation can only get worse if nonzero values of $\kappa$ and $\Gamma$ are allowed.
However, it is possible that a better understanding of these ultracompact alternatives to black holes will uncover constraints that follow from their internal consistency and, in particular, from the laws governing their dynamical evolution (which are largely unknown at the moment). As stressed in Sec. \ref{sec:taxonomy}, most alternative geometries such as the ones of quasi black holes and wormholes are prescribed in static situations. The lack of a framework in which to deal with dynamical processes is highly unsatisfactory, and is arguably the main criticism that can be raised against these models on purely theoretical grounds.

One may expect that it would be difficult to reach model-independent conclusions, given that different models could display very different dynamical behavior. However, it has been shown recently~\cite{Carballo-Rubio2018} that certain model-independent dynamical considerations are restrictive enough to lead to a consistency relation that takes the form of a lower bound on $\mu$. These model-independent considerations reduce essentially to the observation that the boundary (i.e., surface) of standard celestial objects evolves following causal trajectories in spacetime. Note that these trajectories need not correspond to actual moving particles, as this growth will be generally caused by the stacking or piling up of different particles of matter. But this growth is nevertheless driven by physical interactions, which must propagate in a causal manner. On the other hand, trapping horizons are known to be spacelike for standard accreting matter~\cite{Hayward1993,Ashtekar2004}. There is a clear tension between these two different behaviors. This tension results into two possibilities: (i) the CMO is less compact than a given threshold, so that its surface can grow in a timelike (or at most, null) manner without forming trapping horizons, or (ii) the CMO is more compact than this threshold, hence developing trapping horizons in a given interval of time that can be calculated.

Option (i) translates into a consistency constraint that depends on the parameter $\mu$ and the particular model of accreting matter. In spherical symmetry and using the Vaidya geometry, this consistency constraint can be obtained analytically~\cite{Carballo-Rubio2018},
\begin{equation}\label{eq:lower}
\mu\geq\frac{4G\dot{M}}{c^3}.
\end{equation}
Equivalently, we can write this as $\dot M c^2 \leq \mu P_{\rm P}/4$ where $P_{\rm P}=c^5/G$ is the Planck power, which may represent the maximum luminosity attainable in physical processes \cite{Cardoso2018}. In more general situations this consistency constraint would take a different form, displaying for instance additional quantities such as the angular momentum of the CMO or the accreting matter. It would be necessary to extend this simple estimate in order to take into account these effects and obtain more precise constraints, although Eq.~\eqref{eq:lower} can be used in order to extract some conclusions that are unlikely to be changed by these additional considerations. For instance, we can evaluate the lower bound above for Sgr A* using $\dot{M}\gtrsim 10^{-11}\ M_\odot\mbox{ yr}^{-1}$, which yields
\begin{equation}
\mu\geq\mathscr{O}(10^{-24}).
\end{equation}
Recalling Eq. \eqref{eq:muquad}, this value corresponds to
\begin{equation}
\ell \gtrsim 1\mbox{ cm}.
\end{equation}
This lower bound is strong enough in order to show for instance that quasi black holes with $\ell\sim\ell_{\rm P}$, which have values of $\mu$ more than 60 orders of magnitude smaller, must develop trapping horizons during their lifetime. Let us stress that models for the formation of quasi black holes through short-lived bouncing geometries involve the formation of trapping horizons for finite periods of time, that can be as short as $\tau\sim \tau^{(1)}$.

Contrary to the upper bounds analyzed above, Eq.~\eqref{eq:lower} is not affected by the phenomenological parameters $\kappa$ and $\Gamma$. Since it boils down to a statement about causality, the argument behind Eq.~\eqref{eq:lower} depends only on the location of the region in which the interactions between the accreting matter and the CMO take place, and not on the particular details of this interaction.\footnote{Of course, this constraint relies on the assumption that the interactions involved are local and causal in nature. This might not be the case if large quantum effects are involved in the stability of quasi black holes given that, e.g.,~all of the energy conditions (including the dominant one) could be violated in these scenarios, although not much more can be said without detailed models.}
The mechanism leading to this lower bound has also implications for gravitational waves, which are discussed in Sec. \ref{sec:echoes}.

\subsubsection{Hunting shadows \label{sec:shadow}}

The last observational channel employing electromagnetic waves that we want to discuss is based on the detection of light that gets as close as possible to the CMO, without being captured by the gravitational field of the latter. From the analysis of null geodesics in Sec. \ref{sec:infalling}, we can infer that the point of no return is determined by the photon sphere. If we imagine a congruence of light rays that are directed towards the CMO from a source that is far away, whether or not these rays are trapped by the gravitational field of the CMO depends on the value of the angular momentum for each geodesic. Null geodesics with $L<L_\star$ are captured, while those with $L>L_\star$ are dispersed due to the angular momentum barrier. Light rays with $L=L_\star$ would follow a circular orbit at $r=r_{\rm ph}$ around the CMO, although this orbit is unstable so that these light rays will eventually be captured or will escape to infinity. Hence, in this spherically symmetric situation, it is the photon sphere that marks the division between these two different behaviors of light rays.

Consequently, the observation of light rays around a black hole should reveal a shadow (or more appropriately a ``silhouette" given that there is no physical surface on which the shadow is cast) that is associated with the photon sphere~\cite{Eisenstaedt1987,Bardeen1973,Luminet1979,Cunha2017,Bisnovatyi-Kogan2017,Cunha2018} (although its size does not directly corresponds to the size of the latter due to lensing effects, and is weakly sensitive on its relative distance from the horizon). This particular observable has received widespread attention in the astrophysics community, and it is often described as ``imaging the event horizon of black holes''~\cite{Falcke1999}. However, as we have emphasized, the length scale that controls this phenomenon is not the Schwarzschild radius $r_{\rm s}$, but rather the much larger $r_{\rm ph}=3r_{\rm s}/2$. The gap between these two distances is macroscopic, $r_{\rm ph}-r_{\rm s}=r_{\rm s}/2$. In other words, any object that is compact enough to have a photon sphere ($R<r_{\rm ph}$) should display the necessary physical characteristics to lead to a similar (and in most cases, indistinguishable) shadow. This has been recently stressed by several authors~\cite{Vincent2015,Cardoso2016b,Lamy2018,Cunha2018b,Shaikh2018b}. However, certain analyses of particular models of gravastars~\cite{Sakai2014} and wormholes~\cite{Shaikh2018} suggest the existence of peculiar characteristics that might allow distinguishing these alternatives from black holes in future experiments, although a more systematic studies of these claims would be desirable.

On general grounds, we can make use of our phenomenological parametrization in order to highlight the difficulties of using the observation of black hole shadows as a tool to probe the intrinsic properties of astrophysical black holes. Let us consider for instance an ultracompact object with $R=r_{\rm s}(1+\Delta)>r_{\rm s}$ that has a negligible reflection coefficient, $\Gamma=0$, for electromagnetic waves. As discussed in Sec. \ref{sec:supacc} the light that is trapped inside the photon sphere, that would disappear down the Schwarzschild radius for black holes, can be emitted from the surface of the object after being absorbed. This will lead to a very faint (for $\Delta\ll1$) emission of electromagnetic radiation that will be superimposed to the shadow that can be calculated in classical general relativity. Hence, the only way to rule out this kind of situation using solely the observation of the shadow would be being able to discard the existence of this faint emission in the dark region of the shadow. It seems however difficult to attain the precision needed to obtain competitive constraints on the value of $\Delta$, in comparison for instance with the ones obtained for Sgr A* in Sec. \ref{sec:supacc}. Moreover, it is important to keep in mind that these constraints would always take the form of upper bounds on this quantity. Overall, we can conclude that the existence of horizons cannot be decided on the basis of the observations of the shadows of astrophysical black holes \emph{only}. However, new experimental efforts such as the Event Horizon Telescope (EHT) or \mbox{BlackHoleCam} (e.g., \cite{Goddi2017}) will certainly improve our understanding of the environment of Sgr A* and will allow more accurate measurements of its accretion rate \cite{Ricarte2014,Pu2018}, which would increase the accuracy of the constraints that follow from other observational channels and which have been described above in this section. The EHT may also be able to place constraints on possible deviations from the Kerr geometry \cite{Younsi2016} and on the strength of soft fluctuations of the geometry around black holes \cite{Giddings2014,Giddings2016,Giddings2017} (and perhaps other scenarios that include long-range modifications \cite{Haggard2016}). Within out parametrization, we can describe the latter models in terms of tails $\epsilon(t,r)$ with compact support.

\subsubsection{Bursts \label{sec:bursts}}

The search for EM bursts has been claimed in the past as a possible strategy for detecting the outcome of bouncing geometries. Of course, if the previously discussed instability of regular black holes implies their conversion into bouncing solutions, the same observations will be relevant for constraining them. It is less clear if the possible conversion of regular black holes or short-lived bounces into quasi black holes and wormholes would imply any transient burst and what it might depend on. It is natural to conjecture that, if these objects are the outcomes of a series of rapid bounces with short timescale $\tau^{(1)}$, high-energy quasi-periodic bursts  with typical frequency $1/\tau^{(1)}$ should be expected \cite{Barcelo2014}. However, without detailed models that describe, for instance, the damping of these oscillations, not much more can be said at the moment.

While what we said above holds in the case of short-lived bounces of typical timescale $\tau^{(1)}$, more complex is the case of long-lived bounces with $\tau=\tau^{(2)}$, for which several phenomenological studies have been performed in the literature \cite{Barrau2014,Barrau2015,Barrau2016,Vidotto2018}. In this case no relevant signal is expected up to  this timescale while two distinct components are predicted as being associated to the typical size of the exploding object (infrared component) and to the typical energy of the universe at the moment of its formation (ultraviolet component). If $\tau=\tau^{(2)}$, it is reasonably arguable that only primordial black holes, which formed in the early universe, would have the size and the lifetime for exploding soon enough so that we could observe the corresponding signals.

For primordial black holes whose lifetime is of the order of the Hubble time, it was shown that the infrared component of the signal could get up to the GeV scale and be peaked in the MeV, while the ultraviolet part of the burst is expected to be in the TeV range~\cite{Barrau2016,Vidotto2018}. If confirmed by more accurate modelling, this would place the search for the bursts associated to bouncing geometries within the realm of current high energy astrophysics experiments (provided that a sufficient number of primordial black holes is created in the early universe). Furthermore, the fact that bursts further away in redshift would correspond to less massive and more primordial objects implies that their higher peak frequency would partially compensate their higher cosmological redshift~\cite{Vidotto2018}. This is a peculiar behaviour that might be used a a signature for this kind of signals and might help distinguish them from other, more standard, astrophysical emissions.

With regards to our parameterization, it is clear that the detection of one of these bursts could be used to cast constraints on both $\tau_{\pm}$ (depending on the particular scenario), but would not tell us much about tails, $\epsilon(t,r)$.

\subsection{Gravitational waves}

The detection of gravitational waves in LIGO and VIRGO~\cite{Abbott2016,Abbott2016b,Abbott2017,Abbott2017b,Abbott2017c,TheLIGOScientific2017} opens up additional possibilities for testing the properties of astrophysical black holes. One of the best sources of gravitational waves are the merger of compact objects: astrophysical black holes~\cite{Abbott2016,Abbott2016b,Abbott2017,Abbott2017b,Abbott2017c} or neutron stars~\cite{TheLIGOScientific2017}. The merger of compact objects releases abundant information about their nature, although part of it is difficult to extract due to the intrinsically nonlinear nature of the process. It is also interesting that, as discussed below, observations using electromagnetic and gravitational wave observations are complementary, in the sense that the different nature of the physical processes involving these forms of radiation makes the corresponding observational channels more sensible to (and therefore more suitable to measure) different phenomenological parameters. We will illustrate this point using the parametrization introduced in Sec. \ref{sec:param}.

\subsubsection{Coalescence of compact objects \label{sec:merger}}

The waveform produced in a merger can be roughly divided into three main parts: (i) the inspiral phase in which the two objects are far apart, (ii) the merging phase in which the two objects enter in direct contact, and (iii) the ringdown phase that describes the relaxation of the outcome of the merging phase. These three phases are defined by the different physical processes taking place; from a mathematical perspective, these phases are also characterized by the different techniques that are most appropriate for extracting the corresponding gravitational wave signatures.

\begin{itemize}
\item{Inspiral phase: In the inspiral phase the two objects are far apart, so that Newtonian gravity can be used in order to describe this phase to a good approximation. Hence, as in the discussion of orbiting stars in Sec. \ref{sec:orbits}, the details of the near-horizon geometry will not appreciably affect the evolution of the system in this phase. However, while in this phase, binary systems could still possibly display detectable differences with respect to a binary of black holes if the two objects have surfaces instead of horizons (i.e., if $\Delta>0$), due to the effects that the gravitational field of each of the two objects can have on the internal structure of its companion through the induced tidal forces. All known results regarding classical general relativity black holes are consistent with these objects having identically zero tidal deformability~\cite{poisson2014gravity,Binnington2009,Gurlebeck2015}. However, ultracompact configurations without horizons can be tidally deformed~\cite{Cardoso2017b}. Moreover, if $\Gamma\neq0$ the object would decrease its tidal heating (measured in terms of the amount of gravitational radiation that the object absorbs~\cite{Hartle1973,Hughes2001}). It has been argued~\cite{Maselli2018} (see also \cite{Johnson-McDaniel2018}) that both effects could be used in order to place upper bounds on the values of these two parameters $\Delta$ and $\Gamma$ using data from the Laser Interferometer Space Antenna (LISA), with constraints on the value of $\Gamma$ being the most promising ones. Soft fluctuations of the near-horizon geometry, given by tails $\epsilon(t,r)$ with compact support, can be also constrained using this part of the waveform \cite{Liebling2017}.}
 
\item{Merger: The dynamics in the merger phase is highly nonlinear, which renders most of the parameters in our phenomenological parametrization in Sec. \ref{sec:param} useless. In fact,  virtually nothing is known about this nonlinear regime in theories beyond general relativity. It seems that this problem has to be addressed numerically and on a case-by-case basis. However, some of the models discussed in Sec. \ref{sec:taxonomy} could leave an imprint during this phase. In particular, short-lived bouncing geometries would disrupt the merger on timescales of the order of $\tau_-$, perhaps leading to the formation of quasi black holes as proposed in~\cite{Barcelo2014,Barcelo2016}. This may create a distinctive periodic pattern~\cite{Barcelo2014,Barcelo2015} that is similar to the (linear) phenomenon of gravitational wave echoes discussed in the next section, sharing the typical values of the timescale between subsequent echoes but being inherently nonlinear. For completeness, we also mention that horizonless configurations may also lead to electromagnetic afterglows in this phase~\cite{Abramowicz2016,Vachaspati2016}. However, the strength of this emission is largely unknown, and also it is not clear how such a phenomenon would avoid being suppressed by the lensing discussed in Sec. \ref{sec:infalling}.}

\item{Ringdown: The ringdown part of the signal can again be described making used of a linear analysis, in terms of the so-called quasinormal modes~\cite{Cardoso2003,Konoplya2011}. The corresponding waveform is typically given by a linear combination of damped sinusoids. Shortly after the first detection of gravitational waves, a theoretical analysis~\cite{Cardoso2016a} demonstrated explicitly that the form of this part of the signal is associated with the photon sphere at $r_{\rm ph}=3r_{\rm s}/2$, and not the horizon. Hence, a similar comment as in Sec. \ref{sec:shadow} applies: testing the damping of the waveform does not allow drawing certain conclusions about the near-horizon geometry, such as the existence of horizons. It is also worth noting that there are geometries with different $\epsilon(t,r)$ that are still compatible with the detected signals \cite{Konoplya2016}. However, as discussed in the next section, modifications of the near-horizon geometry may trigger new characteristic effects in the late-time ringdown.}
\end{itemize}

\subsubsection{Echoes in the late-time ringdown \label{sec:echoes}}

After the relaxation of the object produced in the merger through the emission of gravitational waves, its properties could still leave imprints in the late-time gravitational-wave signal. These imprints would be the result of the interaction of the gravitational radiation, emitted previously, with the central object. As discussed in detail below, for sufficiently compact situations a significant fraction of this gravitational radiation is backscattered by the gravitational field of the central object, traveling back to and interacting with the latter. The phenomenological parameter that control this late-time behavior is the reflection coefficient $\Gamma$. For a black hole, $\Gamma=0$, which means that all backscattered radiation will disappear down its gravitational well. If $\Gamma\neq0$, some of this radiation would bounce back from the object and could be measured by distant detectors.

Let us start with the simplest possible description of the main physics involved, adding additional details progressively. The first element that must be discussed is the mechanism that leads to the backscattering of the initially outgoing radiation~\cite{Cardoso2016}. This can be introduced by considering the propagation of test particles or waves in the Schwarzschild geometry. For instance, the modes of a scalar field $\Phi(t,r,\theta,\varphi)$, in the usual decomposition in spherical harmonics
\begin{equation}
\Phi(t,r,\theta,\varphi)=\sum_{l=0}^{\infty}\sum_{m=-l}^l\frac{\phi_{lm}(t,r)}{r}Y_{lm}(\theta,\varphi),
\end{equation}
satisfy the wave equation
\begin{equation}
\left(-\frac{\partial^2}{\partial t^2}+\frac{\partial}{\partial r_*^2}-V_l\right)\phi_{lm}(t,r)=0,
\end{equation}
where $r_*$ is the standard tortoise coordinate and
\begin{equation}\label{eq:schpot}
V_l=\left(1-\frac{2M}{r}\right)\left(\frac{l(l+1)}{r^2}+\frac{2M}{r^3}\right)
\end{equation}
is the Regge--Wheeler potential. This potential has a maximum in the vicinity of $r_{\rm ph}=3M$, with the deviation from this value controlled by $1/l$, and smaller the larger the value of $l$. The radius $r_{\rm ph}=3M$ marks also the location of the photon sphere (or light ring), namely the innermost circular null geodesic that is stable (as discussed in Sec. \ref{sec:infalling}). Due to the existence of this maximum in the potential, outgoing waves originated at $r<r_{\rm ph}$ will be backscattered. The fraction of backscattered radiation can be calculated explicitly~\cite{Mark2017,Correia2018,Testa2018,Burgess2018}. Only objects that are compact enough to have a photon sphere will display this phenomenon, and therefore we focus in the following on these objects.

In the case of a black hole, the backscattered waves will be lost into the horizon. However, for objects in which $\Gamma\neq0$ and $\Delta\neq0$, part of the incoming radiation will be be reflected outwards. When crossing the photon sphere at $r_{\rm ph}$, part of this radiation will escape and part will be backscattered. This leads to a periodic phenomenon that would produce a series of ``echoes'' of the first event. Slightly modifying Eq. \eqref{eq:tbounce}, the characteristic timescale of this phenomenon is given by
\begin{equation}
T_{\rm echo}=2M-4M\ln(2\Delta)+T_{\rm int},\label{eq:echotime1}
\end{equation} 
where the first two terms on the right-hand side measure the time that a pointlike particle following a radial null geodesic takes to travel from $r_{\rm ph}=3M$ and $R=r_{\rm s}(1+\Delta)$ and then from $R=r_{\rm s}(1+\Delta)$ to $r_{\rm ph}=3M$, and $T_{\rm int}$ provides a measure of the time that the gravitational wave spends inside the central object (that is, in the region $r\leq R$). 

The precise value of $T_{\rm int}$ depends on the particular model being used but, if one ignores the interaction between gravitational waves and the central object (which, as explained below, is most likely \emph{not} consistent for compact enough configurations), this quantity is expected to be of the order of the light-crossing time or, equivalently, proportional to $M$ (an explicit calculation is provided for instance in~\cite{Holdom2016}). Then, for $\Delta\ll1$ the leading order in Eq.~\eqref{eq:echotime1} would be
\begin{equation}
T_{\rm echo}\simeq-4M\ln(\Delta).\label{eq:echotime2}
\end{equation} 
This logarithmic behavior has been already discussed in Sec. \ref{sec:supacc}. The amplitude of these echoes is proportional to the reflection coefficient $\Gamma$ and also depends on the details of the barrier peaked around the photon sphere. Of course, the amplitude of subsequent echoes decreases monotonically, and a power-law for this decay has been found~\cite{Correia2018,Wang2018}.

The reader may have noticed that there seems to be an inconsistency between our treatment of electromagnetic waves in Sec. \ref{sec:infalling}, and the treatment of gravitational waves in this section. More specifically, we are not bringing up the lensing that had to be taken into account in order to describe the behavior of electromagnetic waves. In other words, we are implying that gravitational waves are not affected by this lensing. There are several aspects behind this assumption. The first one is that the processes involving electromagnetic waves have been assumed to be deeply inelastic, following our intuition about the interaction of light and matter in other systems. However, in the linear approximation used to describe gravitational wave echoes, it is assumed that gravitational waves interact elastically with the central object (we critically revise this assumption at the end of this section). If we accept this main difference it follows then, as long as we are analyzing waves with angular momentum below the critical value $L_\star$ derived in Sec. \ref{sec:infalling}, that electromagnetic waves experience lensing effects while gravitational waves are unaffected.

Moreover, it is worth stressing that these two kinds of radiation have different wavelength, which determines whether or not the geometric optics approximation is reasonable. The gravitational waves produced in the merger of two compact objects into a central object of mass $M$ have wavelengths that are comparable to the Schwarzschild radius of the central object \cite{Cutler2002} (of course, there is a distribution in wavelengths, or frequencies, around this typical scale). On the other hand, the electromagnetic waves relevant for our discussion in Sec.~\ref{sec:infalling} have much shorter wavelengths. Hence, it is reasonable to describe the behavior of electromagnetic waves in these spacetimes within the geometric optics approximation, in which the strong lensing of lightlike geodesics by the gravitational field of the central object is unavoidable. As the gravitational waves of the wavelengths involved in the merger cannot be described in this approximation, they may circumvent the attraction of the central object and escape outwards even if having an angular momentum greater than $L_\star$, depending on the value of $l$ (something that would lead to a grey-body feature of the gravitational wave spectrum). We think that this aspect is worth studying in detail.

Previous works (including~\cite{Cardoso2016,Correia2018,Testa2018,Oshita2018,Wang2018}) have not analyzed this issue explicitly, perhaps due to particular choices of initial conditions for gravitational radiation. In fact, it may be the case that phenomenologically reasonable values of $L$ are below $L_\star$, hence preventing lensing to play any role in realistic scenarios. However, in order to deal with realistic situations one would also need to include the angular momentum of the central object, which would certainly change the value of $L_\star$.

On the other hand, let us recall that Eqs.~\eqref{eq:echotime1} and \eqref{eq:echotime2} are strictly valid for null geodesics. Hence, it is assumed in the literature that the geometric optics approximation is indeed reasonable at least for the analysis of certain aspects of the problem, namely the evaluation of the characteristic timescale \eqref{eq:echotime2}. The critical wavelength below which the geometric optics approximation can be used in order to describe the behavior of waves in the potential \eqref{eq:schpot} is substantially larger than $M$ for $\Delta\ll1$ small enough; in fact, this critical wavelength is roughly given by Eq.~\eqref{eq:echotime2}. This restricts the wavelengths for which the discussion of the echoes provided above is consistent:
\begin{equation}\label{eq:filter}
\lambda\ll |\ln\Delta|\times\mathscr{O}(M).
\end{equation}
It is important to analyze the physics associated with this upper bound in more detail, in order to understand for instance how sharp it is. This is moreover relevant for modeling purposes, as Eq.~\eqref{eq:filter} points out that the frequency content of the originally outgoing gravitational radiation will effectively experience a band-pass filter that selects the frequencies that would appear in the subsequent echoes. Moreover, waves with $L>L_\star$ may also experience a lower bound given by $M\lesssim \lambda$, although this is far from clear.

It is illustrative to use our parametrization in Sec. \ref{sec:param} in order to understand the kind of information that can be extracted from the search of echoes in gravitational wave events. The amplitude of gravitational wave echoes would be, following the discussion above, proportional to $\Gamma$. Hence, both the observation and non-observation of echoes can put constraints on the value of this parameter (this is, for instance, the main result in~\cite{Du2018,Barausse2018}). The non-observation of echoes can only constrain this parameter and cannot say anything about the radius $R$ or, alternatively, $\Delta$. Of course, a positive detection of echoes could be used in order to determine the size of the central object, through the use of Eq.~\eqref{eq:echotime2}. The other two parameters which are relevant for the process are $\tau^+$, which has to be greater than the characteristic timescale of echoes (this would place a very uninteresting lower bound on this quantity), and $\tau_-$ which has to be smaller (the consequences of this for theoretical models were analyzed in~\cite{Barcelo2017}).

The interest in this phenomenon has grown after tentative evidence for their existence in LIGO data was claimed~\cite{Abedi2016,Abedi2017}. These works assume crude templates that are missing some of the details in the discussion above and in later works such as~\cite{Correia2018,Wang2018}, the importance of which for data analysis is not yet clear. Moreover, these claims are still controversial, although parts of the initial results have been corroborated by other groups~\cite{Ashton2016,Westerweck2017,Conklin2017,Abedi2018b} (see also \cite{Maselli2017}). For completeness, let us also mention that qualitatively similar claims have been made about the binary neutron star merger GW170817~\cite{Abedi2018}. However, the latter analysis does not make any specific assumptions about the waveform of the echoes and just looks for periodicities. This opens the possibility of alternative explanations for these periodicities, as mentioned in the conclusions of~\cite{Abedi2018} but also explored for instance in~\cite{Pani2018}, and in Sec. \ref{sec:merger} above in which it was stressed that short-lived bouncing geometries are also expected to lead to periodic patterns in the late-time part of gravitational wave signals.

Before ending this section, we want to stress that the discussion above neglects the (generally nonlinear) interaction between gravitational waves and the central object. In practical terms, the echo timescale is calculated in an approximation in which gravitational waves propagate in a fixed background, and the amplitude of the echoes is just proportional to the reflection coefficient $\Gamma$. This issue has been ignored in the literature, but here we want to highlight that this does not seem consistent and that this feature has fundamental implications for the modeling of echoes.

\begin{figure}[h]%
\begin{center}
\vbox{\includegraphics[width=0.3\textwidth]{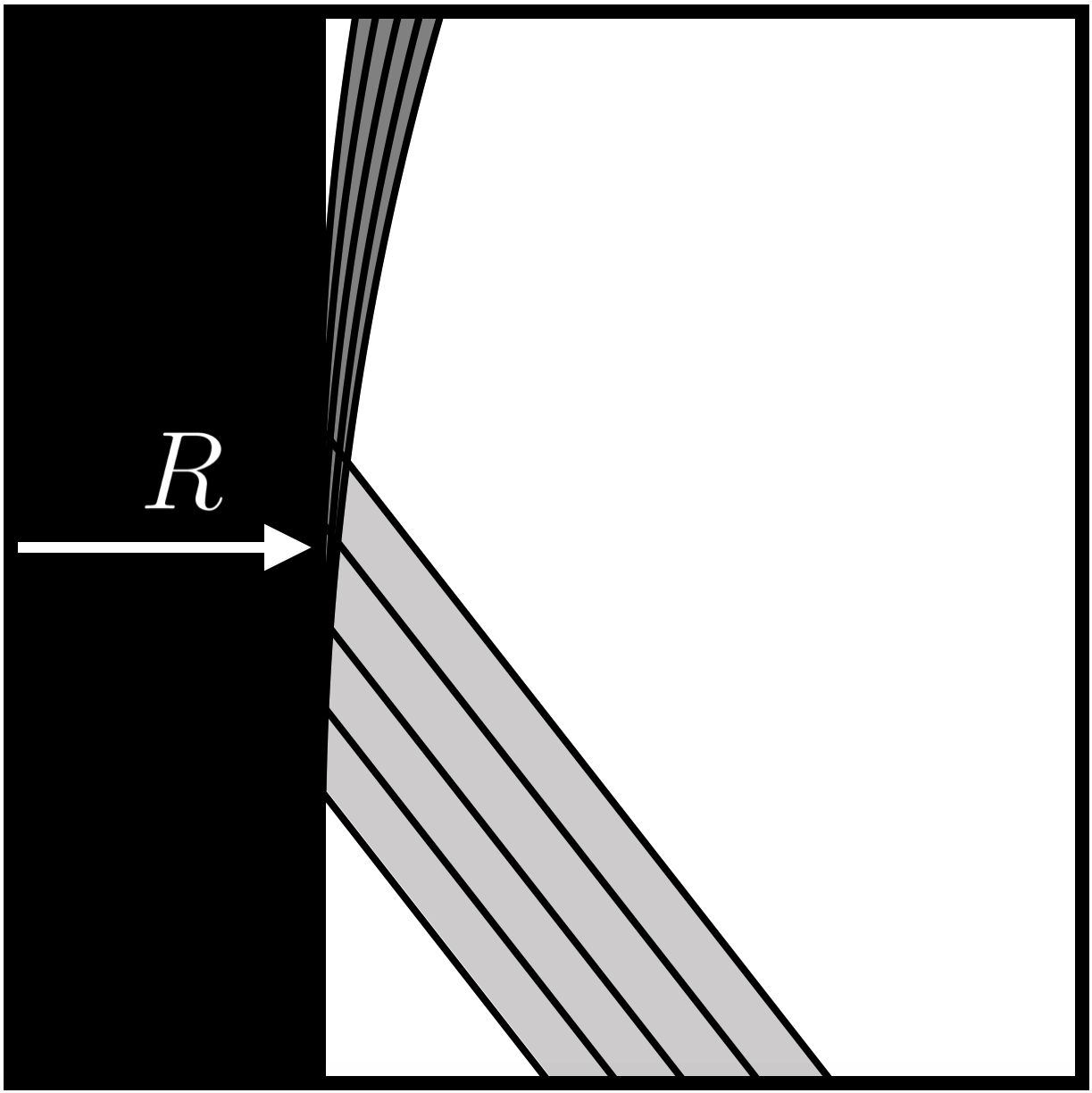}}
\bigskip%
\caption{{Exponential peeling of outgoing geodesics reflected at $r=R$. Even if the ingoing distribution of energy has low density (light gray region), the accumulation of geodesics around the gravitational radius produces high densities (dark gray region) that result in large backreaction effects on the background geometry.}}
\label{fig:peeling}
\end{center}
\end{figure}

Let us start by considering a toy model in the purely classical framework of general relativity, consisting of a perfectly reflecting ($\Gamma=1$) and spherically symmetric mirror with radius $R=r_{\rm s}(1+\Delta)$ enclosing a mass $M$, so that the geometry outside the mirror is Schwarzschild. We now consider an ingoing spherical shell of gravitational radiation, of which we just need to monitor the energy density;  so that we will describe it in terms of pressureless null dust with uniform energy density. This ingoing radiation will be reflected by the mirror and will therefore travel outwards after interacting with the latter. However, the peeling of outgoing null geodesics leads to an accumulation of energy around the gravitational radius (see Fig. \ref{fig:peeling}) For $\Delta\ll1$ this accumulation of energy leads to the formation of trapping (and, in this classical setting, event) horizons even for an extremely modest amount of energy being received and reflected at the mirror~\cite{Carballo-Rubio2018}. Above a certain threshold in the power stored in the gravitational radiation, a black hole will form around the mirror and no radiation will escape to infinity.

The formation of a black hole in this toy model is intimately associated with the breakdown of the linear approximation for the gravitational waves propagating in the background geometry produced by the mirror, given that in the linear approximation these waves will always escape to infinity. Hence, this shows that one has to be careful when using the linear approximation to extract the features of echoes. The formation of a trapping horizon might be avoided if the nonlinear interactions between the ingoing gravitational waves and the central objects are considered. A model-independent outcome of these interactions has to be the expansion of the central object in order to avoid the formation of trapping horizons. This expansion of the object needs energy, which can only be taken from the gravitational radiation. A straightforward application of the argument in~\cite{Carballo-Rubio2018} shows that the more compact the central object is, the larger is the fraction of the energy stored in the gravitational waves that has to be transferred through nonlinear interactions.\footnote{It is important to stress that it is clear that these interactions must be sufficiently exotic (see, e.g.,~\cite{Mathur2017} for a particular discussion) to avoid the formation of trapping horizons (in particular, it seems that these interactions must involve some kind of non-locality), although we will not insist on this point.} If most of the ingoing gravitational waves must interact and transfer their energy to the central object, it is likely that the reflection coefficient $\Gamma$ will be extremely small, and therefore that there would be no echoes. It might be possible that the nonlinear interactions with the central object are elastic and that the energy inside the central object is transferred back to the outgoing gravitational waves after their travel through its interior, although this possibility seems unlikely from a physical perspective. In any case, these arguments show that previous theoretical analysis of this phenomenon are missing important details of the physics involved, which must be incorporated in order to arrive to a consistent picture (and, in particular, to determine whether the existence of echoes really is a robust theoretical prediction, as well as, in case of positive detection, to relate them to the internal details of the central object).

\begin{widetext}
\begin{center}
\begin{table}[h]
\begin{center}
\begin{tabular}{||c||c|c|c|c|c|c||}
\hline
\hline
Model & Stars (EM) & Accretion (EM) & Shadows (EM) & Bursts (EM) & Coalescence (GW) & Echoes (GW) \\
\hline
\hline
Regular black hole & $\epsilon(t,r)$ & \xmark & \xmark & If bouncing  & \xmark & \xmark \\ 
\hline
 Bouncing geometries & $\epsilon(t,r)$ & \xmark &  \xmark & \cmark &  \multicolumn{2}{c}{$\tau_-$ (short-lived)} \vline\,\vline \\
\hline
 Quasi black hole & \xmark & $\mu,\Gamma,\kappa$ & \xmark & \xmark & ${\tau_-},\mu,\Gamma$ & $\Gamma,{[\mu]}$  \\
\hline
 Wormhole &\xmark & \xmark $\ (\Gamma+\kappa=1)$ & \xmark & \xmark & ${\tau_-,\Gamma}$ & $\Gamma,{[\mu]}$  \\
\hline   
\hline
\end{tabular}
\caption{Parameters that can be possibly constrained or measured for the different classes of quantum-modified black holes and different observational channels. {Square brackets are used in order to stress that $\mu$ can be measured using gravitational wave echoes only in the event of a positive detection (in other words, the non-observation of echoes places constraints on $\Gamma$ only).}}\label{tab:2}
\end{center}
\end{table}
\end{center}
\end{widetext}

\section{Conclusion}

In this paper we have studied and parametrized the possible theoretical alternatives to classical black holes and we have discussed the current status of the relevant observational constraints. We have classified the different alternatives into four classes: regular black holes, bouncing geometries, quasi black holes and wormholes, and we have provided a set of phenomenological parameters that identify the key properties of each class. Both electromagnetic and gravitational wave observations can be used in order to constrain these parameters. In Table \ref{tab:2} we have summarized the parameters that can be measured or constrained with each of the different observational channels discussed in this paper.

The most promising observational channel using electromagnetic waves can only probe quasi black holes. Eq.~\eqref{eq:acccons2} represents the most stringent bound that electromagnetic observations can put on a combination of the parameters $\kappa$ and $ \mu$ (a similar relation including $\Gamma$ can be derived). This provides an upper bound on the allowed values of $\mu$ that is generally weaker than the upper bound \eqref{eq:acccons} that does not take into account the absorption coefficient $\kappa$ (as in previous works in the subject). These constraints can become very weak for reasonable values of the parameters involved. We have also reviewed in Sec. \ref{sec:conscon} the existence of lower bounds on the parameter $\mu$ that can also be inferred from the observation of accretion disks around supermassive black holes. These lower bounds provide the most restrictive constraints on quasi black holes and are insensitive to the parameters $\kappa$ and $\Gamma$ (albeit they do rely on the assumption of standard local interactions). Let us stress that all the constraints described in this paper must be taken as order-of-magnitude estimates, as we have not included explicitly the effects of rotation or realistic models of accretion disks, for instance. These additional aspects must be analyzed for each of these different observational channels in order to tighten the accuracy of the corresponding constraints.

As with wormholes, testing regular black holes or long-lived bouncing geometries with electromagnetic observations seems hopeless during most of their extended lifetimes. The final stages in the evolution of these objects may be typically violent, which could lead to prompt emissions of electromagnetic radiation. Hence, quasi black holes are arguably the most interesting scenarios from the perspective of electromagnetic observations, as they offer a number of phenomenological opportunities during all stages of their life cycle (that may involve transients characterized by short-lived bounces).

Regarding gravitational waves, we can conclude that the most promising theoretical scenarios from an observational perspective are quasi black holes, wormholes, and short-lived bouncing geometries. The remaining theoretical scenarios, such as regular black holes, will be very difficult (if not impossible) to probe observationally in the near future, except perhaps for cataclysmic events that may lead to bursts of gravitational radiation (that can be associated, for instance, with long-lived bouncing geometries).

For these theoretical models, observational channels based on gravitational waves are mostly sensitive to the reflection coefficient $\Gamma$. This is due to the main difference with respect to electromagnetic radiation that is typically assumed: gravitational waves interact extremely weakly with standard matter. However, the lack of detailed knowledge of both the matter forming these objects and the possible nonlinear interactions of their gravitational fields with gravitational waves makes it impossible to assume at the moment that these objects will display an appreciable reflection coefficient. In any case, gravitational wave observations are starting to place constraints on this coefficient. These constraints will improve in the near future, thus providing valuable feedback for theoretical research. It is worth stressing that observational channels involving gravitational and electromagnetic waves are therefore complementary, thus providing a strong motivation for a multi-messenger approach to the problem. 

\vspace{10pt}
\acknowledgments

RCR would like to thank Niayesh Afshordi for useful discussions and is also grateful for the hospitality of Perimeter Institute. This research was supported in part by Perimeter Institute for Theoretical Physics. Research at Perimeter Institute is supported by the Government of Canada through the Department of Innovation, Science and Economic Development and by the Province of Ontario through the Ministry of Research and Innovation. MV was supported by the Marsden Fund, which is administered by the Royal Society of New Zealand. MV would like to thank SISSA and INFN (Trieste) for hospitality during the early phase of this work.

\bibliography{refs}	

\end{document}